\documentclass[twocolumn,prb,color,psfig,superscriptaddress]{revtex4}
\usepackage{graphicx}
\usepackage{epsfig}

\def\beq{\begin{equation}}
\def\eeq{\end{equation}}

\begin{document}

\title{Disproportionation and electronic phase separation  in parent manganite LaMnO$_3$}
\author{A.S. Moskvin}
\affiliation{Ural State University, 620083 Ekaterinburg,  Russia}
\date{\today}

\begin{abstract}
Nominally pure undoped  parent manganite LaMnO$_3$ exhibits a puzzling behavior inconsistent with a simple picture of an A-type antiferromagnetic insulator (A-AFI) with a cooperative Jahn-Teller ordering.  We do assign its anomalous properties to charge transfer instabilities and competition between insulating A-AFI phase and metallic-like dynamically disproportionated phase formally separated by a first-order phase transition at T$_{disp}$\,=\,T$_{JT}$$\approx$\,750\,K.  The unconventional high-temperature phase is addressed to be a specific electron-hole Bose liquid (EHBL) rather than a simple "chemically" disproportionated R(Mn$^{2+}$Mn$^{4+})$O$_3$ phase. New phase does nucleate as a result of the charge transfer (CT) instability and evolves from the self-trapped CT excitons, or specific EH-dimers, which seem to be a precursor of both insulating and metallic-like ferromagnetic phases observed in manganites. 
We arrive at highly frustrated system of triplet $(e_g^2){}^3A_{2g}$ bosons moving in a lattice formed by hole Mn$^{4+}$ centers. 
Starting with different experimental data we have reproduced a typical  temperature dependence of the volume fraction of high-temperature mixed-valent EHBL phase. We argue that a slight nonisovalent substitution, photo-irradiation, external pressure or magnetic field gives rise to an electronic phase separation with a nucleation or an overgrowth of EH-droplets. Such a scenario provides a comprehensive explanation of numerous puzzling  properties observed  in parent and nonisovalently doped  manganite LaMnO$_3$ including an intriguing manifestation of superconducting fluctuations.
\end{abstract}
\maketitle


\maketitle

\section{Introduction}
Perovskite manganites RMnO$_3$ (R= rare earth or yttrium) manifest many  extraordinary physical properties. Undoped TbMnO$_3$ and DyMnO$_3$ reveal multiferroic behavior\,\cite{Kimura}. Under nonisovalent substitution all the orthorhombic manganites reveal an insulator-to-metal (IM) transition and colossal magnetoresistance (CMR) effect which are currently explained in terms of an electronic phase separation (EPS) triggered by a hole doping. Overview of the current state of the art  with theoretical and experimental situation in doped CMR manganites R$_{1-x}$Sr(Ca)$_x$MnO$_3$ can be found in  many review articles \cite{Dagotto,Gorkov,Tokura,Dorr,Ramakrishnan}.

However, even nominally pure undoped stoichiometric parent manganite LaMnO$_3$ does exhibit a puzzling behavior inconsistent with a simple picture of an A-type antiferromagnetic insulator (A-AFI) which it is usually assigned to\,\cite{Dagotto,Gorkov,Tokura,Dorr,Ramakrishnan}. First it concerns anomalous transport properties\,\cite{Raffaelle,Hundley,Good1}, photoinduced absorption\,\cite{Mih},  pressure-induced effects\,\cite{Loa}, dielectric anomalies\,\cite{Mondal}, and the high field-induced IM transition\,\cite{Kudasov}. 
Below, in the paper we demonstrate that the unconventional behavior of parent manganite LaMnO$_3$ can be explained to be a result  of an electronic phase separation inherent even for nominally pure stoichiometric manganite with a coexistence of conventional A-AFI phase and unconventional electron-hole Bose liquid (EHBL) which nucleation is a result of a charge transfer (CT) instability of A-AFI phase. 
In a sense, hereafter we report a comprehensive elaboration of a so-called "disproportionation" scenario in manganites which was addressed earlier by many authors, however, by now it was not properly developed. 
 
The paper is organized as follows. In Sec.II we discuss an unconventional first order  phase transition in parent manganite LaMnO$_3$ and argue that it should be addressed to be 
a disproportionation rather than a Jahn-Teller phase transition. Then we show that the resonant X-ray scattering data can be used to  reconstruct a  "phase diagram" which shows a tentative temperature dependence of the volume fraction of two competing phases for parent LaMnO$_3$. The electron-lattice relaxation effects and the self-trapping of the CT excitons with nucleation of electron-hole droplets are considered in Sec.III. In Sec.IV we describe the details of the charge and spin structure of electron-hole dimers to be main building blocks of novel phase in a parent manganite. The effective Hamiltonian of the EHBL phase equivalent to a triplet boson double exchange model is addressed in Sec.V. Numerous optical, magnetic, and other manifestations of the EH dimers and EH droplets in parent and low-hole-doped manganites are considered in Sec.VI. Short comments on  the hole doping effects are made in Sec.VII.  Short conclusions are presented in Sec.VIII. 

\section{First experimental signatures of disproportionation and electronic phase separation in parent manganite L\lowercase{a}M\lowercase{n}O$_3$}

\subsection{Unconventional first order  phase transition in L\lowercase{a}M\lowercase{n}O$_3$}

 Measurements on single crystals of the high-temperature transport and magnetic  properties,\,\cite{Raffaelle,Hundley,Good1,Good2} resonant X-ray scattering\,\cite{Murakami-98,Zimmermann}  and neutron-diffraction\,\cite{Rodriguez} studies    of the RMnO$_3$ family point to a first order $electronic$ phase transition at T=T$_{JT}$ (T$_{JT}$$\approx$\,750\,K in LaMnO$_3$) from the low-temperature orbitally ordered (OO) antiferromagnetic insulating   phase (O' orthorhombic Pbnm) with a cooperative Jahn-Teller ordering of the occupied orbitals of the MnO$_6$ octahedra to a high-temperature charge and orbitally disordered  phase (O orthorhombic, or "pseudocubic" Pbnm). It is worth noting that the "first-orderness" is rather unexpected point for the 
cooperative Jahn-Teller ordering as a common viewpoint implies it is to be a second-order "order-disorder" type  phase transition.
According to the conventional model of the first order phase transitions there are two characteristic temperatures:  T$_1^*$\,$<$T$_{JT}$ and T$_2^{*}$$>$T$_{JT}$ ("supercooling" and "superheating spinodals", respectively) which determine the temperature range of the coexistence of the both phases. The both temperatures are hardly defined for parent manganites.  A change in slope of the temperature dependence of the  thermoelectric
power at T$_1^*$$\approx$\,600\,K in LaMnO$_3$\,\cite{Good1,Good2} is considered to be due to nucleation of an orbitally disordered phase on heating or   homogeneous nucleation of the low-T  OO phase on cooling.  The volume fraction of charge and orbitally disordered  phase is to monotonically grew with increasing temperature in the interval T$_1^*$$<$T$<$T$_{JT}$, but  increase discontinuously on heating across T$_{JT}$. The low-T OO phase looses stability only at T$_2^{*}$$>$T$_{JT}$. Weak diffuse X-ray scattering consistent with orbital fluctuations, was observed in LaMnO$_3$ with the intensity falling gradually with increasing temperature and disappearing above T$_2^{*}$$\sim$\,1000\,K concomitant with the suppression of the octahedral tilt ordering  and  a structural transition to a rhombohedral phase\,\cite{Zimmermann}.
  
The X-ray diffraction  data\,\cite{Prado} for LaMnO$_3$ have revealed a coexistence of two orthorhombic Pbnm phases O' and O in a wide temperature  range both below and above T$_{JT}$. It means that a sizeable volume fraction of large ($\sim$\,1000\,{\AA}) domains of low-(high-) temperature phase survives between T$_{JT}$ and T$^*_2$(T$^*_1$  and T$_{JT}$), respectively. However, it does not prevent the nanoscopic size droplets to survive outside this temperature range. Furthermore, the neutron diffraction measurements (T\,$<$\,300\,K) for several samples of nominal composition LaMnO$_3$ after different heat treatment seemingly provoking the nucleation of a high-temperature phase \,\cite{Huang} have revealed a coexistence of bare orthorhombic A-AFI phase with another orthorhombic and rhombohedral, however, ferromagnetic phases with a considerably ($\sim 2\%$) smaller unit-cell volume and ordering temperatures T$_C$ near T$_N$. Puzzlingly, this coexistence spread out over all temperature range studied from room temperature up to 10\,K.  Similar effects have been observed in a complex (ac initial magnetic susceptibility, magnetization, magnetoresistance, and neutron diffraction) study (T\,$<$\,300\,K) of slightly nonstoichiometric LaMnO$_{3+\delta}$ system\,\cite{Ritter}. Interestingly that all over the ferromagnetic phases the thermal
factors of oxygen atoms present an excess  $\Delta B\sim$\,0.3-0.5\,{\AA}$^2$ as compared with antiferromagnetic A-AFI phase that points to a specific role of dynamic lattice effects. 

  Even in the absence of chemical doping, LaMnO$_3$
shows the ability to accommodate a so-called "oxidative nonstoichiometry", which also involves the partial oxidation of some Mn$^{3+}$ to Mn$^{4+}$ which smaller size leads to an increase of the tolerance factor, thus stabilizing the perovskite structure\,\cite{Alonso}.
 The manganite  crystals grown by the floating zone method seem to preserve well-developed traces of the high-temperature  phase.
Interestingly, that  the LaMnO$_3$ crystals do not tolerate repeated excursions to high temperatures, 800 K, before changing their properties. Such an anomalous memory effect with an overall loss of long-range orbital order in one sample of the LaMnO$_3$ after extended cycling above 1000\,K 
and cooling back to room temperature was  observed by Zimmermann {\it et al.}\,\cite{Zimmermann}. 
It is worth mentioning that the characteristic temperatures T$_1^{*}$, T$_{JT}$, T$_2^{*}$ for the phase transition are believed to depend on the initial content of Mn$^{4+}$\,\cite{Rodriguez}: the sample used in Ref.\,\onlinecite{Norby}  gave T$_{JT}$\,=\,600\,K and T$_2^{*}$\,=\,800\,K,  suggesting the presence of a non-negligble amount of Mn$^{4+}$ that reduces the temperatures of the  phase transition.
 All these data evidence an existence of electronic phase separation inherent for parent stoichiometric LaMnO$_3$ with the phase volume fraction  sensitive to sample stoichiometry, prehistory, and morphology. 
 


\subsection{The disproportionation rather than the JT nature of the phase transition  in parent LaMnO$_3$}

The electronic state in the high-temperature O orthorhombic phase of parent LaMnO$_3$ remains poorly understood. The transport measurements\,\cite{Good1} (resistivity $\rho (T)$ and  thermoelectric power $\alpha (T)$, see, also Refs.\onlinecite{Raffaelle,Hundley,Ahlgren,Buch}) were interpreted by the authors as  a striking evidence of the R(Mn$^{2+}$Mn$^{4+})$O$_3$ disproportionation  rather than a simple orbitally disordered RMn$^{3+}$O$_3$ character of the high-temperature phase. Let us shortly overview the argumentation by Zhou and Goodenough\,\cite{Good1}.   Thermoelectric power reveals an  irreversible change from $\alpha (300\,K)=-\,600\,\mu V/K$ to
about  $550\,\mu V/K$ on thermal cycling to 1100\,K with a nearly zero value at T$>$T$_{JT}$. Small-polaron conduction by a single charge carrier would give a temperature-independent thermoelectric power dominated by the statistical term:
\begin{equation}
\alpha =-(k/e)\ln[(1-c)/c]\, ,
\label{1}
\end{equation}
where c is the fraction of Mn sites occupied by a charge carrier and the spin degree of freedom is lifted by the strong intra-atomic exchange. Near stoichiometry, two types of
charge carriers may be present, but with only one dominating at room temperature to give a large negative or large positive  $\alpha (300\,K)$  for a small value of $c$.   From Eq.\,(\ref{1}) value of $\alpha (300\,K)\approx \pm \,600\,\mu V/K$  in the virgin crystal reflects a small fraction ($c\approx  0.001$) of a disbalance between electron- and hole-like  mobile/immobile charges.  
 An abrupt drop in $\alpha (T)$ and $\rho (T)$ at   T$_{JT}$ to a nearly temperature-independent and a nearly zero value  for T$>$T$_{JT}$  with a  reversible behavior of both  quantities agrees with a phase transition to a fully disproportionated Mn$^{2+}$+Mn$^{4+}$, or, more precisely, to an electron-hole  liquid phase\,\cite{Moskvin-98,Moskvin-02,Moskvin-LTP} with a two-particle transport and $c_{eff}=0.5$. However, the system retains  a rather high value of resistivity, that is the EH liquid phase manifests a "poor" metal behavior. Strictly speaking, the disproportionation  phase transition at T=T$_{disp}$=T$_{JT}$ is governed firstly by a charge order  rather than the orbital order parameter. In other words, the Jahn-Teller ordering at T=T$_{JT}$ only accompanies the charge ordering at T=T$_{disp}$=T$_{JT}$, hence a simplified  Jahn-Teller picture  does misinterpret  a true sense of the phenomenon.

\begin{figure}[t]
\includegraphics[width=8.5cm,angle=0]{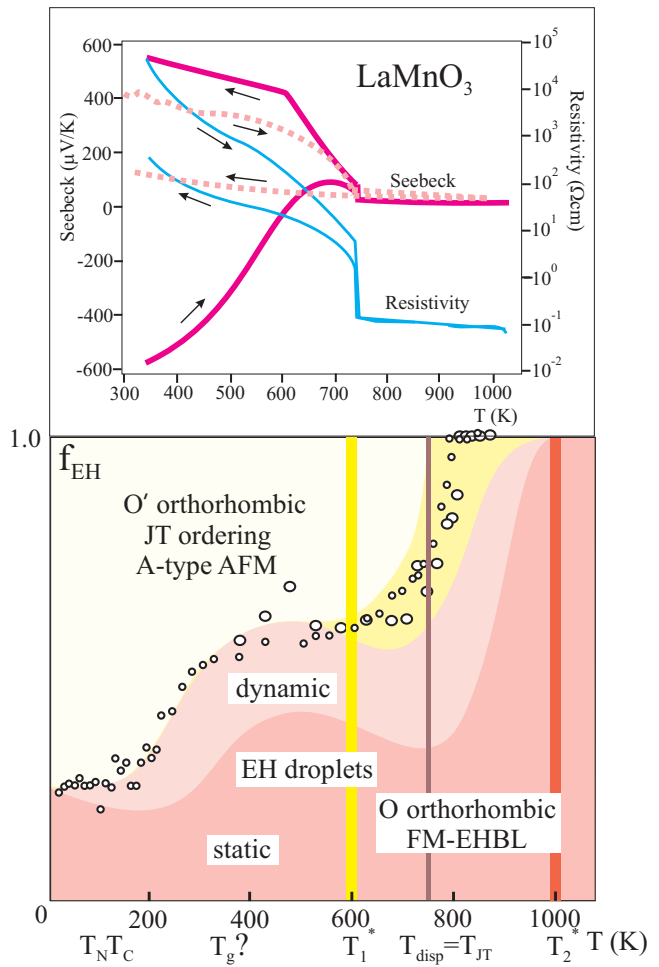}
\caption{(Color online) Top panel: Temperature dependence of thermoelectric power and resistivity in parent manganite LaMnO$_3$ (reproduced from Refs.\,\onlinecite{Good1,Topfer}). Bottom panel: Schematic T-f$_{EH}$ "phase  diagram" of a parent perovskite manganite, f$_{EH}$ being the volume fraction of mixed valence phase. Small and large circles show up experimental data from Refs.\,\onlinecite{Murakami-98,Zimmermann} transformed into a resultant volume fraction of a "non-OO" phase supposed to be a system of static and dynamic EH droplets. Different filling (from top to bottom) points to an A-AFI phase, orbital fluctuation phase near T$_{JT}$, dynamic and static EH droplet phase, respectively. Note a difference in T$_{JT}$ values in Refs.\,\onlinecite{Murakami-98,Zimmermann} and Ref.\,\onlinecite{Good1}} \label{fig1}
\end{figure}

In contrast with the high-temperature  measurements carried out in a vacuum of 10$^{-3}$ 
 torr\,\cite{Good1}, the transport measurements  performed in air\,\cite{Topfer} evidenced another evolution of $\alpha (T)$ (see Fig.\,\ref{fig1}). On heating the thermoelectric power starts from large, but $positive$ values and  on cooling from T$>$T$_{JT}$ $\alpha (T)$ does not return to its original value because the sample, according to authors\,\cite{Topfer}, becomes slightly ($\sim 1 \%$) oxidized. A simple comparison of the two data sets\,\cite{Good1,Topfer} points to an unconventional behavior of parent manganite on crossing the "supercooling spinodal" temperature T$^*_1$. The system can memorize a high-temperature phase up to temperatures below 300\,K. The role of a slight oxidation seemingly reduces to be  an additional regulative factor governing the A-AFI/EHBL phase volume fraction.

 Strong and irreversible temperature dependence of $\alpha (T)$ and $\rho (T)$ at T$<$T$^*_1$ agrees with a scenario of a well developed electronic phase separation with a puzzling electron-hole symmetry and  a strong sensitivity of transport properties both to a sample morphology and quality.  
 The magnitude of the resistivity and character of irreversibility agrees with a  {\it poor metal} like conductivity of high-temperature phase and points to a considerable volume fraction of this phase to survive up to room temperature. 
  Resistivity of different samples of the nominally same composition can differ by orders of magnitude. Interestingly that these data point to a possibility of a colossal, up to six orders of magnitude, variations of resistivity in parent LaMnO$_3$ at a constant temperature well below T$_{JT}$ only due to the variation in its A-AFI/EHBL volume fraction composition which can be realized by the temperature change, pressure, isotopic substitution, application of external magnetic/electric field, and photoirradiation. This behavior can hardly be directly related with the colossal magnetoresistivity observed for the hole doped manganites, however, this novel phase can be an important participant of electronic transformations in manganites.

Below T=T$_1^*$ (T$_1^*$$\approx$\,600\,K
 in LaMnO$_3$), or the temperature of the homogeneous nucleation of the low-T  OO phase, the high-temperature  mixed-valence EH phase loses stability, however, survives due to various charge inhomogeneities forming EH droplets pinned by statically fluctuating electric fields. 
 
\subsection{Temperature dependence of the EHBL volume fraction}  
 
 By now we have no information about how the both phases share the volume fraction on cooling from high temperatures. Clearly, such an information depends strongly on the techniques used. For instance, the both long-lived static domains and short-lived dynamic fluctuations of either phase contribute to optical response, while only the large static domains are seen in conventional x-ray  or neutron scattering measurements. Fortunately, the resonant X-ray scattering data\,\cite{Murakami-98,Zimmermann} can be used to  reconstruct the tentative T-f$_{EH}$ "phase diagram" of a manganite with f$_{EH}$ being a volume fraction of EH droplets. Indeed, the intensity of this scattering depends on the size of the splitting $\Delta$ of the Mn 4p levels, induced by the orbital ordering of Mn 3d$e_g$ states, hence is nonzero only for orbitally ordered Mn$^{3+}$ ions in distorted MnO$_6$ octahedra. The first-order nature of the cooperative JT phase transition in LaMnO$_3$\,\cite{Good1} implies that the local orbital order parameter such as $\Delta$ in Ref.\onlinecite{Murakami-98} remains nearly constant below the transition temperature\,\cite{Sartbaeva}, hence the  temperature behavior of resonant X-ray scattering intensity has to reflect the temperature change of the net (static+dynamic) OO phase volume fraction rather than $\Delta (T)$ effect. This suggestion agrees with the neutron-diffraction studies  by Rodr\'iguez-Carvajal {\it et al.}\,\cite{Rodriguez} evidencing no visible effect of the antiferromagnetic spin ordering at T=T$_N$\,$\approx$\,140\,K on the OO parameter, while the X-ray scattering intensity dramatically (up to 40\%) falls upon heating above $T_N$\,\cite{Murakami-98}. Overall, the temperature dependence of the resonant X-ray scattering intensity in LaMnO$_3$ shows up an unusual behaviour with an arrest or even clear hole  between 300-500 K, a sharp downfall above T=T$^*_1\approx 600$\,K and vanishing right after T=T$_{JT}\approx 750$\,K.  Thus, the X-ray data\,\cite{Murakami-98,Zimmermann}  can be used to find the temperature behavior of the resultant static and dynamic  EH droplet volume fraction in the sample. 

In Fig.\,\ref{fig1} we have reproduced experimental data from Refs.\,\onlinecite{Murakami-98,Zimmermann} renormalized and  transformed into a relative volume fraction of a "non-OO" phase supposed to be an EH droplet phase.  The renormalization implied the low-temperature 75\% volume fraction of the OO phase. 
Different filling (from top to bottom) points to an A-AFI phase, orbital fluctuation phase near T$_{JT}$, dynamic and static EH droplet phase, respectively. Despite the overall fall of the EH droplet volume fraction on cooling from T$_{JT}$, we expect some intervals of the reentrant behavior   due to a subtle competition of two phases. It is clear that any ordering does lower the free energy of the phase thus resulting in a rise of its volume fraction. Taking into account experimental data from Ref.\,\onlinecite{Huang} pointing to  close temperatures of AFI and FI ordering in competing phases (T$_N$ and T$_C$, respectively), we may assign a signature of a reentrant behavior at 400-600\,K to a  glass-like  transition within the EH liquid near T=T$_g$\,$\sim$\,400\,K.
Surely, we are aware that the picture shown in Fig.\,\ref{fig1} is not a real phase diagram, however, it is very  instructive for a qualitative understanding of a complex phase competition in parent manganite.

Concluding the section, we should once more emphasize a dramatic charge instability of parent manganite LaMnO$_3$ with extreme sensitivity to different external factors, sample stoichiometry and prehistory. In this connection, it is worth noting that highly stoichiometric LaMnO$_3$ samples measured by Sub\'{i}as {\it et al.}\cite{Subias-07} did not show noticeable temperature dependence of the resonant intensity for the  (3,0,0)  reflection from 10 to 300\,K, in contrast with the data by Murakami {\it et al.}\cite{Murakami-98}. 
Further work at an even higher temperature range and for different samples seems to be necessary in order to distinctly reveal and examine the phase separated state in a parent manganite.

\section{Electron-lattice relaxation and nucleation of EH droplets in a parent manganite}
\subsection{Electron-lattice relaxation and self-trapping of CT excitons}
 
At first glance the disproportionation in manganese compounds is hardly possible since manganese atom does not manifest a valence-skipping phenomenon as, e.g. bismuth atom which can be found as Bi$^{3+}$ or Bi$^{5+}$, but not Bi$^{4+}$, with a generic bismuth oxide BaBiO$_3$ to be a well-known example of a charge disproportionated system. Strictly speaking, sometimes manganese reveals a valence preference, e.g., while the both Mn$^{2+}$ and  Mn$^{4+}$ are observed in MgO:Mn and CaO:Mn, the Mn$^{3+}$ center is missing\,\cite{MgO}.
Furthermore, the d$^4$  configuration of Mn$^{3+}$ ion is argued\cite{Katayama} to be a missing oxidation state due to the large exchange-correlation energy gain that stabilizes the d$^5$ electronic configuration thus resulting in the charge disproportionation or dynamical charge fluctuation 2\,d$^4$$\rightarrow$d$^3$ + d$^5$.

The reason for valence skipping or valence preference observed for many elements still remains a mistery. Recently, Harrison\,\cite{Harrison} has argued that most likely traditional lattice relaxation effects, rather than any intra-atomic mechanisms (specific behavior of ionization energies, stability of closed shells, strong screening of the high-charged states) are a driving force for disproportionation with formation of so-called "negative-U" centers.

Anyhow the disproportionation in an insulator signals a well developed CT instability.
What is a microscopic origin of the CT instability in parent manganites? The disproportionation reaction can be considered to be a final stage of  a self-trapping of the $d-d$ CT excitons (Mott-Hubbard excitons) that determine the main low-energy CT band peaked near 2 eV in LaMnO$_3$\,\cite{Kovaleva}. Indeed, these two-center excitations due to a charge transfer between two MnO$_6$ octahedra may be considered as quanta of the disproportionation reaction
\begin{equation}
\mbox{MnO}_{6}^{9-}+\mbox{MnO}_{6}^{9-}\rightarrow
\mbox{MnO}_{6}^{8-}+\mbox{MnO}_{6}^{10-}\, ,
\label{d}
\end{equation}
with the creation of electron  MnO$_{6}^{10-}$ and hole MnO$_{4}^{8-}$ centers.
Within a simplest model\,\cite{Moskvin-02} the former corresponds to a nominal $3d^5$  (Mn$^{2+}$) configuration,  while the latter does to the $3d^3$  (Mn$^{4+}$) one.

The minimal energy cost of the optically excited
disproportionation or electron-hole formation in insulating manganites   is
$2.0-2.5$ eV\,\cite{Kovaleva}. However, the question arises, what is the energy cost for the thermal excitation of such a local disproportionation or effective correlation energy $U$?  The answer implies
first of all the knowledge of relaxation energy, or  the energy gain due to the lattice polarization by the
localized charges. The full polarization energy $R$  includes the cumulative
effect of $electronic$ and $ionic$ terms, related with the displacement of
electron shells and ionic cores, respectively\,\cite{Shluger}. The former term $R_{opt}$ is due
to the {\it non-retarded} effect of the electronic polarization by the momentarily
localized electron-hole pair given the ionic cores fixed at their perfect crystal positions.
Such a situation is typical for lattice response accompanying the Franck-Condon
transitions (optical excitation, photoionization). On the other hand, all the
long-lived excitations, i.e., all the intrinsic thermally activated states and
the extrinsic particles produced as a result of doping, injection or optical
pumping should be regarded as stationary states of a system with a deformed
lattice structure. 
\begin{figure}[h]
\includegraphics[width=.48\textwidth]{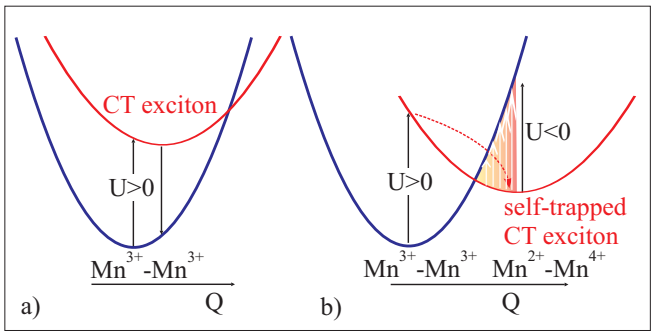}
\caption{(Color online) 
 Simple illustration of the electron-lattice polarization effects for CT excitons (see text for details.)}
\label{fig2}
\end{figure} 

 The lattice
relaxation energies, $-\Delta R_{th}$,  associated with the hole/electron
localisation in 3d oxides are particularly large.  For instance, in LaMnO$_3$ the optical
(non-relaxed) energies of the creation of the hole on Mn and O sites are 2.6
 and 4.9\,eV, respectively, while   $-\Delta R_{th}^{Mn}$=0.7-0.8 and
$-\Delta R_{th}^{O}=2.4$ eV\,\cite{Kovaleva-JETP}. In other words,  the electronic hole is marginally more stable at the Mn-site than at the O-site in the LaMnO$_3$ lattice, however, both possibilities should be treated seriously. 

Shell-model estimations\,\cite{Kovaleva-JETP} yield for the energy of the optically excited
disproportionation (\ref{d}) or electron-hole formation in parent manganite LaMnO$_3$:$E_{opt}\approx 3.7$\,eV, while the respective thermal relaxation energy is estimated as $-\Delta R_{th}\approx 1.0$\,eV. Despite the estimations imply the noninteracting electron and hole centers these are believed to provide a sound background for any reasonable models of self-trapped $d-d$ CT excitons.
Thorough calculation of the localisation energy for electron-hole dimers remains a challenging task for future studies. It is worth noting that despite their very large, several eV magnitudes, the relaxation  effects are not incorporated into current theoretical models of manganites.

Fig.\,2 illustrates two possible ways  the electron-lattice polarization governs the CT exciton evolution. Shown are the adiabatic potentials (AP) for the two-center ground state $M^0-M^0$ configuration and  excited $M^{\pm}-M^{\mp}$ CT, or disproportionated configuration.
The $Q$ coordinate  is related with a lattice degree of freedom.
  For lower branch of AP in the system we have either a single minimum point for the GS configuration (Fig.\,2a) or  a two-well structure with an additional local minimum point (Fig.\,2b) associated with the self-trapped CT exciton. This   "bistability" effect is of primary importance for our analysis. Indeed, these two minima are related with two (meta)stable charge states with and without CT, respectively,  which form two candidates to struggle for a ground state.  It is worth noting that the self-trapped CT exciton may be descibed as a configuration with  negative disproportionation energy $U$. 
 Thus one  concludes that all the systems such as manganites may be divided into two classes: {\it CT stable systems} with the only lower AP branch  minimum for a certain  charge configuration, and bistable, or {\it CT unstable systems} with two lower AP branch minima for two local charge configurations one of which is associated with the self-trapped CT excitons  resulting from self-consistent charge transfer and electron-lattice relaxation.  Such excitons are often related with the appearance
of the ``negative-$U$'' effect.
 It means that the three types of MnO$_6$ centers  MnO$_{4}^{8,9,10-}$ should be considered in manganites on equal footing\,\cite{Moskvin-02,Moskvin-LTP}.
 
Above we have presented a generalized disproportionation scenario for parent manganites in which an unconventional phase state with a 2Mn$^{3+}$$\rightarrow$Mn$^{2+}$+Mn$^{4+}$ disproportionation nominally within manganese subsystem evolves from EH-dimers, or self-trapped $d-d$ CT excitons. However, such a scenario in parent manganites would compete with another, "asymmetric"  disproportionation scenario
\begin{equation}
	 \mbox{Mn}^{3+}+\mbox{O}^{2-}\rightarrow \mbox{Mn}^{2+}+\mbox{O}^{1-}
	 \label{d1}
\end{equation}
 that evolves from a self-trapping of low-energy $p-d$ CT excitons. Indeed, we should make a remarkable observation, which to the best of our knowledge has not been previously reported for these materials, is that the famous "manganite" 2 eV absorption band has a composite structure being a superposition of a rather broad and intensive CT $d-d$ band and several narrow and relatively weak CT $p-d$ bands\,\cite{Kovaleva,Moskvin-PRB}. A dual nature of the dielectric gap in nominally stoichiometric  parent perovskite manganites RMnO$_3$, being formed by a superposition of forbidden or weak dipole allowed p-d CT transitions and inter-site d-d CT transitions means that these  should rather be sorted neither into the CT insulator nor the Mott-Hubbard insulator in the Zaanen, Sawatzky, Allen\,\cite{ZSA}  scheme.  A detailed analysis of the CT $p-d$ transitions in LaMnO$_3$ has been performed by the present author in Ref.\onlinecite{Moskvin-PRB}. Among the first $p-d$ candidates for a self-trapping we should point to the low-energy CT state 
$((t_{2g}^{3}{}^{4}A_{2g};e_{g}^{2}\,{}^{3}A_{2g};{}^{6}A_{1g});
\underline{t}_{1g}){}^{5,7}T_{1g}$ in MnO$_6^{9-}$ octahedron
which arises as a result of the O2p electron transfer from the  highest in energy nonbonding $t_{1g}$ orbital to the $e_g$ manganese orbital. Simplest view of this exciton implies the oxygen  $t_{1g}$ hole rotating around nominally Mn$^{2+}$ ion with ferro- (${}^{7}T_{1g}$) or antiferro- (${}^{5}T_{1g}$) ordering. It has a number of unconventional properties. First, orbitally degenerated ground  $T_{1g}$ state implies a nonquenched orbital moment and strong magnetic anisotropy. May be more important to say that we deal with a Jahn-Teller center unstable with regard to local distortions. Second, we expect a high-spin S=3 ground state ${}^{7}T_{1g}$ because of usually ferromagnetic $p-d$ exchange coupling. Oxygen holes can form so-called O$^-$ bound small polarons\,\cite{Schirmer}.

Shell-model estimations\,\cite{Kovaleva-JETP} yield for the energy of optically excited asymmetric disproportionation (\ref{d1})  in parent manganite LaMnO$_3$:$E_{opt}\approx 4.75$\,eV, while the respective thermal relaxation energy is estimated as $-\Delta R_{th}\approx 1.25$\,eV. However, these qualitative estimations do not concern a number of important points such as p-d and p-p covalency, and a partial delocalisation of oxygen holes.  

A sharp electron-hole asymmetry and a rather big S=3 ground state spin value  most likely exclude the self-trapped $p-d$ CT excitons as candidates to form a high-temperature T\,$>$\,T$_{JT}$ phase of parent manganite. However, the "dangerous" closeness to the ground state makes them to be the potential participants of any perturbations taking place for parent manganites.  
 
\subsection{Nucleation of EH droplets in a parent manganite}

The AP bistability in CT unstable insulators points to tempting perspectives  of their evolution under either external impact.  
Metastable CT excitons in the CT unstable $M^0$ phase, or EH-dimers,  present candidate
 "relaxed excited states" to struggle for stability with ground state and the natural nucleation centers for electron-hole liquid phase. What way the CT unstable $M^0$ phase can be transformed into novel phase?  It seems likely that such a phase transition
could be realized due to a mechanism familiar to semiconductors with filled bands such as Ge and Si where   given  certain conditions one observes a formation of metallic EH-liquid as a result of the exciton decay\,\cite{Rice}.
 However, the  system of strongly correlated electron $M^{-}$ and hole
$M^{+}$ centers  appears to be equivalent to an electron-hole
Bose-liquid in contrast with the electron-hole  Fermi-liquid in
conventional semiconductors. The Mott-Wannier excitons in the latter wide-band systems
dissociate easily producing two-component electron-hole gas or plasma\,\cite{Rice}, while 
 small CT excitons  both free and self-trapped  are likely
   to be stable with regard to  the EH dissociation. At the same time, the two-center
    CT excitons have a very large fluctuating electrical dipole moment
     $|d|\sim 2eR_{MM}$ and can
be involved into attractive electrostatic dipole-dipole interaction. Namely
this is believed to be important incentive to the proliferation of excitons and
its clusterization. 
The CT excitons are proved to attract each other and form molecules
called biexcitons, and more complex clusters, or excitonic strings, where the individuality of the
separate exciton is likely to be lost. Moreover, one may assume that  like the semiconductors with indirect band gap structure, it is energetically favorable for the system to separate into a low density exciton
phase coexisting with the microregions of a high density two-component phase
composed of electron $M^-$ and hole $M^+$ centers, or EH droplets. 
Indeed, the excitons may be considered to be well defined entities only at
small content, whereas at large densities their coupling is screened and their
overlap becomes so considerable  that they loose individuality and we come to
the system of electron $M^-$ and hole $M^+$ centers, which form a metallic-like  electron-hole Bose liquid  with a main two-particle transport mechanism\,\cite{Moskvin-LTP}. 
An increase of
injected excitons in this case merely increases the size of the EH droplets,
without changing the free exciton density. 

An EH droplet seems to have no distinct boundary, most likely it looks like a core with more or less stable electron and hole centers surrounded by a cloud of metastable CT excitons. 
{\it Homogeneous nucleation} implies the spontaneous formation of EH droplets
due to the thermodynamic fluctuations in exciton gas. Generally speaking, such
a state with a nonzero volume fraction of EH droplets and the spontaneous
breaking of translational symmetry can be stable in nominally pure insulating
crystal. However, the level of intrinsic nonstoichiometry in 3d oxides is
significant (one charged defect every 100-1000 molecular units is common). The
charged defect produces random electric field, which can be very large (up to
$10^8$ Vcm$^{-1}$) thus promoting the condensation of CT excitons and the
{\it inhomogeneous nucleation} of EH droplets.

 Deviation from the neutrality  implies the
existence of additional electron, or hole centers that can be the natural
 centers for the  inhomogeneous nucleation of the EH droplets.
Such droplets are believed to provide a more effective screening of the
electrostatic repulsion for additional electron/hole centers, than the parent
insulating phase. As a result, the electron/hole injection to the insulating
$M^0$ phase due to a nonisovalent substitution as in La$_{1-x}$Sr$_x$MnO$_{3}$
 or change in  stoihiometry as in
La$_{x}$MnO$_{3}$, LaMnO$_{3-\delta}$
 or field-effect is believed to shift the phase
equilibrium from the insulating state to the unconventional electron-hole Bose
liquid, or in other words  induce the insulator-to-EHBL phase transition.
 This process results in a relative
increase of the energy of the parent phase and creates proper conditions for
its competing with others phases capable to provide an effective screening of
the charge inhomogeneity potential. The strongly degenerate system of electron
and hole centers in EH droplet is one of the most preferable ones for this
purpose. At the beginning (nucleation regime) an EH droplet nucleates as a
nanoscopic cluster composed of several number of neighboring electron and hole
centers pinned  by disorder potential.  It is clear that such a situation does not exclude the self-doping with the formation of a self-organized collective charge-inhomogeneous state in systems
which are near the charge instability.

EH droplets  can manifest itself remarkably  in various properties of the
3d oxides even  at  small volume fraction, or in a ``pseudo-impurity regime''. Insulators in this regime should be considered as phase inhomogeneous systems with, in general, thermo-activated mobility of the inter-phase
boundaries.  On the one hand, main features of this ``pseudo-impurity regime'' would be determined by the
partial intrinsic contributions of the appropriate phase components with
possible limitations imposed by the finite size effects. On the other hand, the
real properties will be determined  by the peculiar geometrical factors such as
a volume fraction, average size of droplets and its dispersion, the shape and
possible texture of  the droplets, the geometrical relaxation rates. These
factors are tightly coupled, especially near phase transitions for either phase
(long range antiferromagnetic ordering for the parent phase, the charge
ordering and other phase transformations  for the EH droplets) accompanied  by
the variation in a relative volume fraction.  

Numerous
examples of the unconventional behavior of the 3d oxides in the pseudo-impurity
regime could be easily explained with taking into account the inter-phase
boundary effects (coercitivity, the mobility threshold, non-ohmic conductivity, oscillations,
relaxation etc.) and corresponding characteristic quantities.
Under increasing doping the pseudo-impurity regime with a relatively small
volume fraction of EH droplets (nanoscopic phase separation) can gradually transform into a  macro- (chemical) ``phase-separation regime'' with a sizeable volume fraction of EH droplets, and finally to a new EH liquid phase.     
 
\section{Electron-hole dimers in parent manganite}

\subsection{EH-dimers: $physical$ versus $chemical$ view}

 Parent manganites are believed to be unconventional systems which are unstable with
regard to a self-trapping  of the low-energy charge transfer excitons which are precursors of nucleation of the EH Bose liquid.  Hereafter we should emphasize once more that a view of the self-trapped CT exciton to be a  Mn$^{2+}$-Mn$^{4+}$  pair is typical for a $chemical$ view of disproportionation, and is strongly oversimplified.  Actually we deal with an EH-dimer to be a dynamically charge fluctuating system of coupled electron MnO$_{6}^{10-}$ and hole MnO$_{4}^{8-}$ centers having been glued in a lattice due to a strong electron-lattice polarization effects. In other words, we should proceed with a rather complex $physical$ view of disproportionation phenomena which first implies a  charge exchange reaction
\begin{equation}
\mbox{Mn}^{2+}+\mbox{Mn}^{4+} \leftrightarrow \mbox{Mn}^{4+}+\mbox{Mn}^{2+}\, ,	
\end{equation}
governed by a two-particle charge transfer integral
\begin{equation}
t_B=\langle \mbox{Mn}^{2+}\mbox{Mn}^{4+}|\hat H_B|\mbox{Mn}^{4+}\mbox{Mn}^{2+}\rangle \, ,	
\end{equation}
 where $\hat H_B$ is an effective two-particle (bosonic) transfer Hamiltonian, and we assume a parallel orientation of all the spins.
As a result of this quantum process the bare ionic states with site-centred charge order and the same bare energy $E_0$ transform into two EH-dimer states with an indefinite valence and bond-centred charge-order
\begin{equation}
|\pm \rangle =\frac{1}{\sqrt{2}}(|\mbox{Mn}^{2+}\mbox{Mn}^{4+}\rangle \pm |\mbox{Mn}^{4+}\mbox{Mn}^{2+}\rangle )	
\end{equation}
with the energies $E_{\pm}=E_0\pm t_B$. In other words, the exchange reaction restores the bare charge symmetry.
 In both $|\pm\rangle $ states the site manganese valence is indefinite with quantum  fluctuations between +2 and +4, however, with a mean value +3.
Interestingly that, in contrast with the  ionic states, the EH-dimer states $|\pm \rangle $  have both a distinct electron-hole and inversion symmetry, even parity ($s$-type symmetry) for $|+ \rangle $, and odd parity ($p$-type symmetry) for $|- \rangle $ states, respectively. The both states are coupled by a large electric-dipole matrix element:
\begin{equation}
\langle +|\hat {\bf d}|-\rangle =2eR_{MnMn}\, , 	
\end{equation}
where $R_{MnMn}$ is a Mn-Mn separation.
The two-particle transport 
 Mn$^{2+}$-Mn$^{4+}$$\rightarrow$Mn$^{4+}$-Mn$^{2+}$ can be realized through two successive one-particle processes with the $e_g$-electron transfer  as follows
$$
\mbox{Mn}^{2+}+\mbox{Mn}^{4+} \stackrel{e_g}{\rightarrow}  \mbox{Mn}^{3+}+\mbox{Mn}^{3+} \stackrel{e_g}{\rightarrow} \mbox{Mn}^{4+}+\mbox{Mn}^{2+}\, ,
$$
hence the two-particle transfer integral $t_B$ can be evaluated as follows:
\begin{equation}
t_B=-t_{e_g}^2/U \,,	
\end{equation}
where $t_{e_g}$ is one-particle transfer integral for $e_g$ electron, $U$ is a mean transfer energy. It means that  the two-particle bosonic transfer integral  can be directly coupled with the kinetic $e_g$-contribution $J_{kin}^{e_g}$ to Heisenberg exchange integral. Both $t_B$ and $J_{kin}^{e_g}$ are determined by the second order one-particle transfer mechanism.
It should be noted that negative sign of the two-particle CT integral $t_B$ points to the energy stabilization of the $s$-type EH-dimer state $|+ \rangle $.

Second, one should emphasize once more that the stabilization of EH-dimers is provided by a strong electron-lattice effect with a striking intermediate oxygen atom polarization and  displacement concommitant with charge exchange. In a sense, the EH-dimer may be addressed to be a bosonic counterpart of the Zener Mn$^{4+}$-Mn$^{3+}$ polaron\,\cite{Zener}. It is no wonder that even in a generic disproportionated system BaBiO$_3$ instead of simple checkerboard charge ordering of Bi$^{3+}$ and Bi$^{5+}$ ions we arrive at CDW state with the alteration of expanded Bi$^{(4-\rho )+}$O$_6$ and compressed Bi$^{(4+\rho )+}$O$_6$ octahedra with $0<\rho \ll 1$\,\cite{BaBiO-Merz}. Enormously large values of oxygen thermal parameters in BaBiO$_3$\,\cite{Chaillout} evidence a great importance of dynamical oxygen breathing modes providing some sort of a "disproportionation glue". Sharp rise of the oxygen thermal parameter in the high-temperature O phase of LaMnO$_3$\,\cite{Rodriguez} or in several "competing" phases found by Huang {\it et al.}\,\cite{Huang} as compared with the bare AFI phase is believed to be a clear signature of the manganese disproportionation.

The formation of EH-dimers seems to be a more  complex process than it is assumed in simplified approaches such as Peierls-Hubbard model (see, e.g., Ref.\onlinecite{Nasu}) or  Rice-Sneddon model\,\cite{Rice-Sneddon}. As a rule, these focus on the breathing mode for the intermediate oxygen ion and neglect strong effects of the overall electron-lattice relaxation. The EH-dimer can be viewed as an Jahn-Teller center (JT polaron) with the energy spectrum perturbed by strong electron-lattice effects.

Thus we see that a simple chemical view of the disproportionation should be actually replaced by a more realistic physical view that implies a quantum and $dynamical$ nature of the disproportionation reaction.

\subsection{EH-dimers:  spin structure }

Let's apply to spin degrees of freedom which are of great importance for magnetic properties both of isolated EH-dimer and of the EHBL phase  that evolves from the  EH-dimers. The net spin of the EH-dimer is ${\bf S}={\bf S }_1+{\bf S}_2$, where ${\bf S }_1$ ($S_1=5/2$) and ${\bf S }_2$ ($S_1=3/2$) are spins of Mn$^{2+}$ and Mn$^{4+}$ ions, respectively. In nonrelativistic approximation the spin structure of the EH-dimer will be determined by isotropic Heisenberg exchange coupling 
\begin{equation}
V_{ex}=J\,({\bf S }_1\cdot {\bf S }_2),	
\end{equation}
 with $J$ being an exchange integral, and two-particle charge transfer characterized by a respective transfer integral which depend on spin states as follows:
\begin{equation}
\langle \frac{5}{2}\frac{3}{2};SM|\hat H_B|\frac{3}{2}\frac{5}{2};SM\rangle =\frac{1}{20}S(S+1)\,t_B\, ,	
\end{equation}
 where  $t_B$ is a spinless  transfer integral. Making use of this relation we can introduce an effective spin-operator form for the boson transfer as follows:
 \begin{equation}
	\hat H_B^{eff}=\frac{t_B}{20}\left[2(\hat {\bf S}_1\cdot\hat {\bf S}_2)+S_1(S_1+1)+S_2(S_2+1)\right]\, ,
\end{equation}
 which can be a very instructive tool both for qualitative and quantitative analysis of boson transfer effects, in particular, the temperature effects. For instance, the expression points to a strong, almost twofold, suppression of effective transfer integral in paramagnetic phase as compared with its maximal value for a ferromagnetic ordering.

 Both conventional Heisenberg exchange coupling and unconventional two-particle bosonic transfer, or bosonic double exchange can be easily diagonalized in the net spin S representation so that for the energy we arrive at
\begin{equation}
E_S=\frac{J}{2}[S(S+1)-\frac{25}{2}]\pm \frac{1}{20}S(S+1)\,t_B\,,	
\end{equation}
where $\pm$ corresponds to two quantum superpositions $|\pm\rangle $ written in a spin representation as follows
\begin{equation}
|SM\rangle _{\pm} =\frac{1}{\sqrt{2}}(|\frac{5}{2}\frac{3}{2};SM\rangle \pm |\frac{3}{2}\frac{5}{2};SM\rangle )\, ,	
\end{equation}
with $s$- and $p$-type symmetry, respectively.  It is worth  noting that the bosonic double exchange contribution formally corresponds to ferromagnetic exchange coupling with $J_B=-\frac{1}{10}|t_B|$. 

We see that the cumulative effect of the Heisenberg exchange and the bosonic double exchange results in a stabilization of the S\,=\,4 high-spin
 (ferromagnetic) state of the EH-dimer provided $|t_B|>10J$ and  the S\,=\,1 low-spin (ferrimagnetic) state otherwise. Spin states with intermediate S values: S= 2, 3 correspond to a classical noncollinear ordering.
 
 To estimate the both quantities $t_B$ and $J$ we can address the results of a comprehensive analysis of different exchange parameters in 
  perovskites RFeO$_3$, RCrO$_3$, and RFe$_{1-x}$Cr$_x$O$_3$ with Fe$^{3+}$ and Cr$^{3+}$ ions\,\cite{Ovanesyan} isoelectronic with Mn$^{2+}$ and Mn$^{4+}$, respectively. For the superexchange geometry typical for LaMnO$_3$\,\cite{Alonso} with the Mn-O-Mn bond angle  
   $\theta \approx 155^{\circ}$ the authors have found J=J(d$^5$-d$^3$)=\,+7.2\,K while for $J(e_ge_g)\approx -t_B =295.6$\,K. In other words, for a net effective echange integral we come to a rather large value: $J_{eff}=J-0.1|t_B|\approx 22.4$\,K.  Despite the antiferromagnetic sign of the Heisenberg superexchange integral these data unambiguously point to a dominant ferromagnetic contribution of the bosonic double exchange mechanism.    
   
\begin{figure}[t]
\includegraphics[width=8.5cm,angle=0]{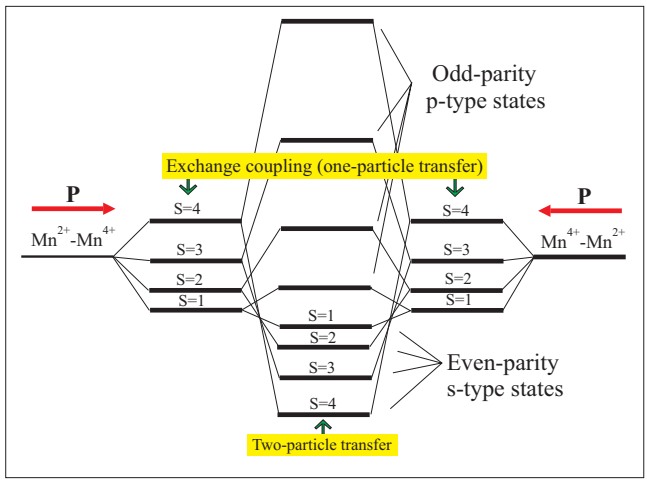}
\caption{(Color online) Spin structure of the self-trapped CT exciton, or EH-dimer with a step-by-step  inclusion of one- and two-particle charge transfer. Arrows point to electric dipole moment for bare site-centred dimer configurations.} \label{fig3}
\end{figure}

    It is worth noting that  the authors\,\cite{Ovanesyan} have predicted the sign change of the superexchange integral in the d$^5$-O$^{2-}$-d$^3$ system Fe$^{3+}$-O$^{2-}$-Cr$^{3+}$ in perovskite lattice from the antiferromagnetic to ferromagnetic one on crossing the superexchange bonding angle  $\theta \approx 162^{\circ}$. Interestingly that the parameter $J(e_ge_g$)\,$\approx - t_B$ is shown to rapidly fall  with the decrease in the bond angle $\theta$ in contrast with  $J=J$(d$^5$-d$^3$) which reveals a rapid rise with $\theta$. For the bond angle $\theta =143^{\circ}$ typical for the heavy rare-earth manganites RMnO$_3$ (R=Dy, Ho, Y, Er)\,\cite{Alonso} the relation between $t_B\approx - 153.8$\,K and $J=J$(d$^5$-d$^3$)\,$\approx 14.4$\,K\,\cite{Ovanesyan} approaches to the critical one: $|t_B|=10\,J$ evidencing a destabilization of the ferromagnetic state for the EH-dimers. 
In other words,  the structural factor plays a significant role for stabilization of one or another spin state of the EH-dimers. Spin structure of the EH-dimer given antiferromagnetic sign of exchange integral J$>$0 and $|t_B|=20\,J$ is shown in Fig.\,\ref{fig3}. We see a dramatic competition of two opposite trends, governed by one- and two-particle transport, respectively.

EH-dimers can manifest typical superparamagnetic behavior with large values of the effective spin magnetic moment up to $\mu_{eff}\approx 9\mu_B$.
Both bare Mn$^{2+}$ and Mn$^{4+}$ constituents of the EH-dimer are S-type ions, i.e. these have an orbitally nondegenerated ground state that predetermines a rather small spin anisotropy.

\begin{figure}[t]
\includegraphics[width=8.5cm,angle=0]{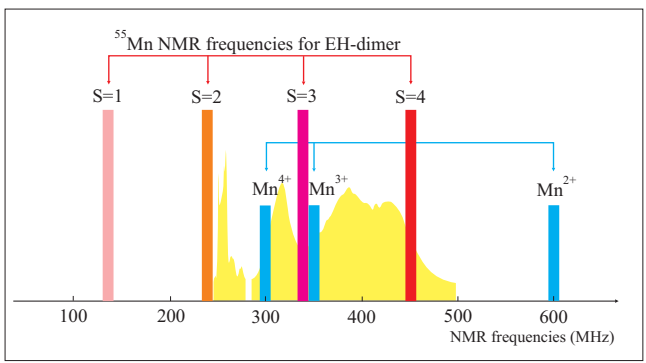}
\caption{(Color online) ${}^{55}$Mn NMR frequencies for bare Mn$^{4+,3+,2+}$ ions in LaMnO$_3$\,\cite{Tomka,Allodi,Shimizu} and theoretical predictions for the EH-dimer in  different spin states. Shown by filling is a ${}^{55}$Mn NMR signal for slightly nonstoichiometric LaMnO$_3$ reproduced from Ref.\,\onlinecite{Kapusta}} \label{fig4}
\end{figure}

Local magnetic fields on the manganese nuclei in the both bond-centred $|SM\rangle _{\pm}$ states of the  EH-dimer are the same and determined as follows:
\begin{equation}
{\bf H}_n=\frac{1}{2}\left[\frac{S(S+1)+5}{2S(S+1)}A_2+\frac{S(S+1)-5}{2S(S+1)}A_4\right]\langle {\bf S}\rangle	\, ,
\end{equation}
where $A_2$, $A_4$ are hyperfine constants for Mn$^{2+}$ and Mn$^{4+}$, respectively, and we neglect the effects of transferred and supertransferred hyperfine interactions. Starting with typical for  Mn$^{2+}$ and Mn$^{4+}$ values of $\frac{5}{2}A_2=600$ MHz and $\frac{3}{2}A_4=300$ MHz, respectively, we arrive at maximal values of ${}^{55}$Mn NMR frequencies  for S=\,4, 3, 2,  1 spin states of the EH-dimer to be 450, 342.5, 237, and 135 MHz, respectively. The ${}^{55}$Mn NMR frequencies for bare Mn$^{4+,3+,2+}$ ions in LaMnO$_3$\,\cite{Tomka,Allodi,Shimizu} and theoretical predictions for the EH-dimer in  different spin states are shown in Fig.\,\ref{fig4}.
Comparing these values with two bare frequencies we see that ${}^{55}$Mn NMR can be an useful tool to study the EH-dimers in a wide range from bond-centred to site-centred states. Experimental  ${}^{55}$Mn NMR signal for slightly nonstoichiometric LaMnO$_3$\,\cite{Kapusta} is shown in Fig.\,\ref{fig4} by filling (see Sec.VI for discussion).

Concluding the subsection we should point to  unconventional magnetoelectric properties of the EH-dimer. Indeed, the two-particle bosonic transport and respective kinetic contribution to stabilization of the ferromagnetic ordering  can be suppressed by a relatively small electric fields that makes the  EH-dimer to be a  promising magnetoelectric cell especially for the heavy rare-earth manganites RMnO$_3$ (R=Dy, Ho, Y, Er) with supposedly  a ferro-antiferro instability. In addition, it is worth noting a strong anisotropy of the dimer's electric polarizability. In an external electric field the EH-dimers tend to align along the field.

\subsection{EH-dimer dynamics. Immobile and mobile dimers.}

 Above we addressed  the
internal electron-hole  motion in a localized immobile EH-dimer resulting in an $s-p$ splitting.  However, the EH-dimer can move in 3D lattice thus developing new translational and rotational modes. For simplicity, hereafter we address an ideal cubic perovskite lattice where the main modes are  rotations of the hole (electron) around the
electron (hole) by $90^{\circ}$ and $180^{\circ}$, and axial translations.
It is interesting to note that the $90^{\circ}$- and $180^{\circ}$-rotation of
the hole(electron) around the electron(hole) corresponds to the 2nd nearest
neighbor $nnn$- and 3rd nearest neighbor $nnnn$-hopping of the hole(electron)
MnO$_{6}^{8-}$(MnO$_{6}^{10-}$) center in the lattice
formed by the MnO$_{6}^{9-}$ centers. 
We can introduce a set of
transfer parameters to describe the dimer dynamics:
$$
t_{s} = - t_{p} \approx \frac{1}{2}(t^{e}_{nnnn}+t^{h}_{nnnn})\, ;
$$
$$
t_{sp}=-t_{ps}\approx  \frac{1}{2}(t^{e}_{nnnn}-t^{h}_{nnnn}) \; ,
$$
for the collinear exciton motion, and
$$
t_{s}^{xy}=- t_{p}^{xy} \approx
 \frac{1}{2}(t^{e}_{nnn}+t^{h}_{nnn}) \, ;
 $$
 $$
t_{sp}^{xy}=t_{ps}^{xy}\approx  \frac{1}{2}(t^{e}_{nnn}-t^{h}_{nnn}) \; 
$$
corresponding to a $90^{\circ}$ rotation ($x\rightarrow y$  motion) of the exciton.
All these parameters have a rather clear physical sense. The electron (hole) transfer integrals for collinear exciton
transfer $t^{e,h}_{nnnn}$ are believed to be smaller  than
$t^{e,h}_{nnn}$ integrals for rectangular transfer. In other words,
the two-center dimers prefer to move "crab-like", rather than in
the usual collinear mode. This implies a large difference for the
dimer dispersion in [100] and [110] directions.

\begin{figure}[t]
\includegraphics[width=8.5cm,angle=0]{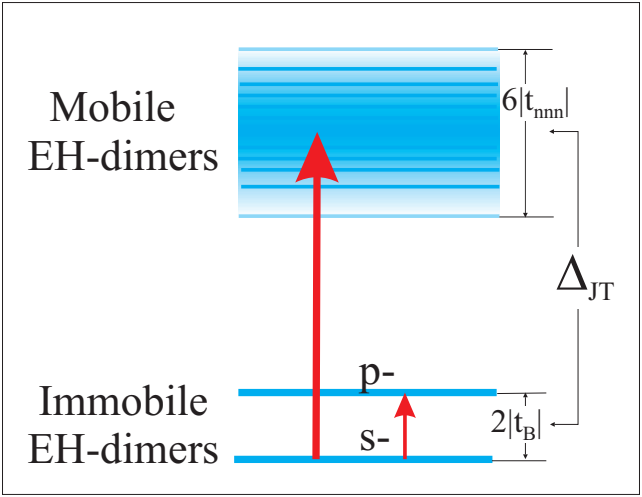}
 \caption{(Color online) Schematic  energy spectrum of immobile (localized)  and mobile EH-dimers. Bold arrows point to  allowed electro-dipole  transitions.}
 \label{fig5}
 \end{figure} 
The motion of the EH-dimer in the bare LaMnO$_3$ lattice with the orbital order of the Jahn-Teller Mn$^{3+}$ ions  bears an activation character with an activation energy $\Delta E=\frac{1}{2}\Delta_{JT}$, where $\Delta_{JT}$ is the Jahn-Teller splitting of the $e_g$ levels in Mn$^{3+}$ ions. 
Thus one may conclude that the EH-dimer energy band in the bare LaMnO$_3$ lattice would be composed of the   low-energy subband of immobile localized EH-dimers, or $sp$-doublet with the energy separation of 2$|t_B|$ and the high-energy subband of mobile EH-dimers shifted by $\frac{1}{2}\Delta_{JT}$ with the bandwidth $W\sim 6t_{nnn}$, where $t_{nnn}$ is an effective next-nearest neighbor $e_g-e_g$ transfer integral in Mn$^{3+}$-Mn$^{3+}$ pair. Schematically the spectrum is shown in Fig.\,\ref{fig5}. An optical portrait of the EH-dimer  bands  is composed of a rather narrow low-energy line due to electro-dipole CT $s-p$ transition for immobile dimers peaked at $E_{SP}=2|t_B|$ and a relatively broad high-energy line due to electro-dipole photo-induced dimer transport peaked at $E\approx \frac{1}{2}\Delta_{JT}+|t_B|$. To estimate these energies one might use our aforementioned estimates for $|t_B|\approx 0.03$\,eV and reasonable estimates of the Jahn-Teller splitting $\Delta_{JT}\approx 0.7$\,eV (see, e.g. Ref.\,\onlinecite{Kovaleva}). Thus we predict a two-peak structure of the EH-dimer optical response with a narrow line at $\sim 0.06$ eV and a broad line at $\sim 0.4$ eV.
Our estimate of the $sp$-separation $E_{SP}=2|t_B|$ does not account for the Jahn-Teller polaronic effects in the EH-dimer that can result in its strong increase.

It is worth noting that the activation character for the motion of the EH-dimer in parent manganite lattice implies the same feature for the generic 2 eV $d-d$ CT exciton resulting in its weak dispersion. Indeed, the resonant inelastic x-ray scattering (RIXS) experiments
on parent manganite LaMnO$_3$ by Inami {\it et al.}\,\cite{Inami} found the energy dispersion of the 2.0-2.5 eV peak to be less than a few hundred meV.

\subsection{EH-dimers: EH dissociation and recombination.}
The EH-dimer dissociation or uncoupling energy may be estimated to be on the order of 1.0-1.5 eV. The EH coupling within the dimer is determined by a cumulative effect of electrostatic attraction and local lattice relaxation (reorganization) energy.

The EH recombination in the EH-dimer resembles an inverse disproportionation  reaction
\begin{equation}
\mbox{MnO}_{6}^{8-}+\mbox{MnO}_{6}^{10-}\rightarrow
\mbox{MnO}_{6}^{9-}+\mbox{MnO}_{6}^{9-}\, .
\label{dd}
\end{equation}
The  inverse counterpart of 2 eV $d$-$d$ CT transition in the bare parent manganite  
is expected to have nearly the same energy. The CT transition (\ref{dd}) in EH-dimer can be induced only in ${\bf E}\parallel {\bf R}_{MnMn}$ polarization.
 However, this CT transition can be hardly photoinduced from the ground $s$-type state of the EH-dimer in contrast with the $p$-type state due to selection rules for electro-dipole transitions. It means that at least at rather low temperatures $kT\ll 2|t_B|$ the EH recombination band would be invisible that is  the optical response of EH-dimers would be reduced to two aforementioned low-energy bands that are developed within the energy gap of the bare parent manganite. In addition, we should point to different $p-d$ CT transitions within electron  $\mbox{MnO}_{6}^{10-}$ and hole $\mbox{MnO}_{6}^{8-}$ centers with the onset energy near 3 eV.

It is worth noting that the overall optical response of the EH-dimers in weakly distorted perovskite lattice is expected to be nearly isotropic at variance with the CT response of parent LaMnO$_3$ in its bare A-AFI phase\,\cite{Kovaleva}.

\section{Electron-hole Bose liquid: the triplet boson double exchange model}

\subsection{Effective Hamiltonian}
 
To describe the electron-hole Bose liquid or EHBL phase  that evolves from EH-dimers we  restrict ourselves with orbital singlets $^{6}A_{1g}$ and $^{4}A_{2g}$ for the electron MnO$_{6}^{10-}$ and hole MnO$_{6}^{8-}$ centers, respectively. Specific electron configurations of these centers, $t_{2g}^3;{}^{4}A_{2g}e_g^2;{}^{3}A_{2g}:{}^{6}A_{1g}$ and $t_{2g}^3;{}^{4}A_{2g}$, respectively enable us to consider the electron center MnO$_{6}^{10-}$ to be composed of the hole MnO$_{6}^{8-}$ center and a two-electron $e_g^2;{}^{3}A_{2g}$ configuration which can be viewed as a composite triplet boson.
In the absence of the external
magnetic field the effective Hamiltonian of the  electron-hole Bose liquid takes the form of the Hamiltonian of the quantum lattice Bose gas of the triplet bosons with an exchange coupling:
$$ \hat H = \hat H_{QLBG}+\hat
H_{ex}=\sum_{i\not=j,m}t_{B}(ij)\hat B_{im}^{\dagger}\hat B_{jm}
$$
$$
+\sum_{i>j}V_{ij}n_{i}n_{j}
-\mu\sum_{i}n_{i} 
+\sum_{i>j}J_{ij}^{hh}(\hat{\bf S}_{i}\cdot\hat{\bf S}_{j}) 
$$
\begin{equation}
+
\sum_{i\not=j}J_{ij}^{hb}(\hat{\bf s}_{i}\cdot\hat{\bf S}_{j})
+\sum_{i>j}J_{ij}^{bb}(\hat{\bf s}_{i}\cdot\hat{\bf s}_{j})
+\sum_{i}J_{ii}^{hb}(\hat{\bf s}_{i}\cdot\hat{\bf S}_{i})\,.
\label{H}
\end{equation}
Here $\hat B_{im}^{\dagger}$ denotes the $S=1$ boson creation operator with a spin projection $m$  at the site $i$; $\hat B_{im}$ is a corresponding annihilation operator. The boson number operator $\hat n_{im}=\hat B_{im}^{\dagger}\hat B_{im}$ at $i$-site
due to the condition of the on-site infinitely large repulsion  $V_{ii}\rightarrow+\infty$ ({\it hardcore boson}) can take values 0 or 1. 

The first term in (\ref{H}) corresponds to the kinetic energy of the bosons,
$t_B(ij)$ is the transfer integral. The second one reflects the effective
repulsion ($V_{ij}>0$) of the bosons located on the neighboring sites. The
chemical potential $\mu$ is introduced to fix the boson concentration:
$
n=\frac{1}{N}\sum_{i}\langle \hat n_{i}\rangle\,.
$
For EHBL phase in parent manganite we arrive at the same number of electron and hole centers, that is to $n=\frac{1}{2}$.
The  remaining terms in  (\ref{H}) represent the Heisenberg exchange
interaction between the spins of the hole centers (term with $J^{hh}$), spins
of the hole centers and the neighbor boson spins (term with $J^{hb}$), boson
spins (term with $J^{bb}$), and the very last term in (\ref{H}) stands for the
intra-center Hund exchange between the boson spin and the spin of the hole center.
In order to account for the Hund rule one should consider $J^{hb}_{ii}$ to be
infinitely large ferromagnetic. 

Generally speaking, this model Hamiltonian describes the system that can be
considered as a Bose-analogue of the {\it one orbital} double-exchange model system\,\cite{Dagotto}.

Aforementioned estimates for different superexchange couplings given the bond geometry typical for LaMnO$_3$ predict antiferromagnetic coupling of the $nn$ hole centers ($J^{hh}>0$), antiferromagnetic coupling of the two nearest neighbor bosons  ($J^{bb}>0$), and ferromagnetic coupling of the boson and the nearest neighbor hole centers ($J^{hb}<0$).
In other words, we arrive at highly frustrated system of triplet bosons moving in a lattice formed by hole centers when the hole centers  tend to order G-type antiferromagnetically, the triplet bosons tend to order ferromagnetically both with respect to its own site and its nearest neighbors. Furthermore, nearest neighboring bosons strongly prefer an antiferromagnetic ordering. Lastly, the boson transport prefers an overall ferromagnetic ordering.

\subsection{Implications for phase states and phase diagram}

By now we have no comprehensive analysis of phase states and phase diagram for the generalized triplet boson double exchange model. The tentative analysis of  the  model in framework of a mean-field approximation (MFA)\,\cite{Avvakumov} allows to predict a very rich phase diagram even  at half-filling ($n=\frac{1}{2}$) with a rather conventional diagonal long-range order (DLRO) with ferromagnetic insulating or ferromagnetic metallic phase, and unconventional off-diagonal long-range order (ODLRO) with a coexistence of superfluidity of triplet bosons and ferromagnetic ordering. However, it is unlikely that the MFA approach can provide a relevant description of such a complex system.

Some implications may be formulated from the comparison with familiar double exchange model\,\cite{Dagotto}, singlet boson Hubbard model (see, e.g.,\,\cite{bubble}), and with generic bismuthate oxide BaBiO$_3$ as a well documented disproportionated system which can be described as a 3D system of the spin-singlet local bosons.

  If the boson transfer to exclude we arrive at a spin system resembling that of mixed orthoferrite-orthochromite LaFe$_{1-x}$Cr$_{x}$O$_3$ ($n_B=\frac{1}{2}(1-x)$) which is an G-type antiferromagnet all over the dilution range 0\,$<x<$\,1 with T$_N$'s shifting from T$_N$=740\,K for LaFeO$_3$ to T$_N$=140\,K for LaCrO$_3$\,\cite{Kadomtseva}. However, at variance with a monovalent (Fe$^{3+}$-Cr$^{3+}$) orthoferrite-orthochromite the Mn$^{2+}$-Mn$^{4+}$ charge system in  the EHBL phase would reveal a trend to a charge ordering, e.g. of a simple checkerboard G-type in LaMnO$_3$ ($n_B=\frac{1}{2}$). It is worth noting that the naively expected large values of a boson-boson repulsion $V_{ij}$ would result in a large temperature T$_{CO}$ of the charge ordering well beyond room temperature. However, the manganites must have a large dielectric function and a strong screening of the repulsion, hence moderate values of $V_{nn}$ and T$_{CO}$'s predicted.   
  
  However, such a scenario breaks when the boson transport is at work. It  does suppress the both types of charge and spin ordering and we arrive most likely at an inhomogeneous system with a glass-like behavior of charge and spin subsystems  which does or does not reveal a long-range ferromagnetic order at low temperatures. A question remains, whether the EHBL Hamiltonian (\ref{H})  can lead to uniform solutions beyond MFA?

  According to experimental data\,\cite{Huang} the novel phases in LaMnO$_3$ which we relate with EHBL, exhibit a long-range ferromagnetic order below T$_C\approx$\,140\,K, however, with rather small values of a mean magnetic moment, that agrees with a spin inhomogeneity. It is worth noting that the glass scenario implies a specific "freezing" temperature T$_g$ to be a remnant of the MFA critical temperature. Such a temperature should be revealed in physical properties of the system.



With a  deviation from half-filling to $n_B<\frac{1}{2}$ the  local triplet bosons gain in freedom to move and improve their kinetic energy. On the other hand it is accompanied by a sharp  decrease in the number of the boson-boson pairs with the most strong $e_g$-$e_g$ antiferromagnetic coupling. In other words, a  ferromagnetic metallic (FM) phase becomes a main candidate to a ground state.

Interestingly, that an intent reader can note that here we describe main features of  phase diagrams typical for hole doped manganites such as La$_{1-x}$Ca$_x$MnO$_3$. Indeed, this resemblance seems not to be accidental one and   points to a profound role of the EHBL phase in unconventional properties of doped manganites as well.

One of the most intriguing and  challenging issues is related with the probable superfluidity of the triplet local bosons. Indeed, the boson transfer integral $t_B$ defines a maximal temperature T$_{max}\approx t_B$ of the onset of local superconducting fluctuations in the hard-core boson systems\,\cite{Micnas}. Our estimations point to T$_{max}$$\approx$\,300-700\,K, where the low bound is taken from theoretical estimations while the  upper bound is derived from optical data on the 0.1 eV spectral feature. However, these high values of T$_{max}$ do not give rise to optimistic expectations regarding the high-T$_c$ bulk superconductivity in the EHBL phase of parent manganites  first because of a spin frustration.  Nevertheless, despite the fact that the emergence of a bulk superconductivity in a highly frustrated multi-component EHBL phase seems to be a very uncommon phenomenon, the well developed local superconducting fluctuations can  strongly influence the transport as well as other physical properties.
A detailed analysis of the bosonic double-exchange model, in particular, of the off-diagonal superconducting order with the superfluidity of the triplet local bosons, remains to be a challenging issue for future studies.
  
It is worth noting that the electron-lattice coupling can be strongly involved into the build-up of the electronic structure of the bosonic double exchange model, in particular, strengthening the EH dimer fluctuations.

\section{Experimental manifestation of EH droplets in parent and low-hole-doped manganites}
Above, in Sec.II we addressed some experimental data that somehow pointed to a disproportionation scenario and have been used to start with a detailed analysis of a novel phase. Hereafter, we address different new experimental data that support our scenario in some details.

\subsection{Optical response of electronically phase separated manganites}

\begin{figure}[t]
\includegraphics[width=7.5cm,angle=0]{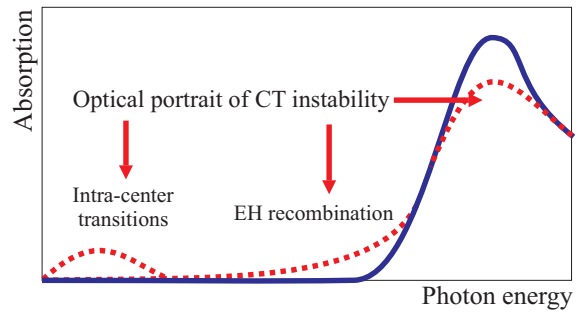}
\caption{(Color online) Optical response (schematically) of the self-trapped CT excitons and EH droplets (dotted curves). Arrows point to a spectral weight transfer from the bare CT band to the CT gap with an appearance of the mid-gap bands and/or smearing of the fundamental absorption edge.} \label{fig6}
\end{figure}

 The CT unstable systems will be characterized by a well developed volume fraction of the short- and long-lived CT excitons or the EH droplets that can give rise to a specific optical response in a wide spectral range due to different \emph{p-d} and \emph{d-d} CT transitions.  First, these are the low-energy  intracenter CT transitions and high-energy inverse \emph{d-d} CT transitions, or EH recombination process in EH-dimers and/or nanoscopic EH centers, and different high-energy CT transitions in electron and hole centers. It is worth noting that, strictly speaking, the optical measurements  should always display a larger volume fraction of EH droplets as compared with static or quasistatic measurements because these "see" short-lived droplets as well.
 What are the main optical signatures of the CT instability? A simplified picture implies  the spectral weight transfer from the bare CT band to the CT gap with an appearance of the mid-gap bands and smearing of the fundamental absorption edge.  Such a transformation of the optical response is shown schematically in  Fig.\,\ref{fig6}. The transferred spectral weight can be easily revealed in the spectral window of the bare insulator to be a direct indicator of the CT instability. 
It is worth noting that the fragile "matrix-droplets" structure of the parent manganites makes the optical response to be very sensitive to such factors as temperature, sample shape (bulk crystal, thin film) and quality, external magnetic field, that can explain some inconsistensies observed by different authors (see, e.g., Refs.\,\onlinecite{Kovaleva,Tobe,Kim2,optics-PS}). Great care is needed, if one wants to separate off the volume fraction effects to obtain the temperature behavior of spectral weight for  certain band and compare the results with those observed by different groups on different samples.

 \begin{figure}[b]
\includegraphics[width=7.5cm,angle=0]{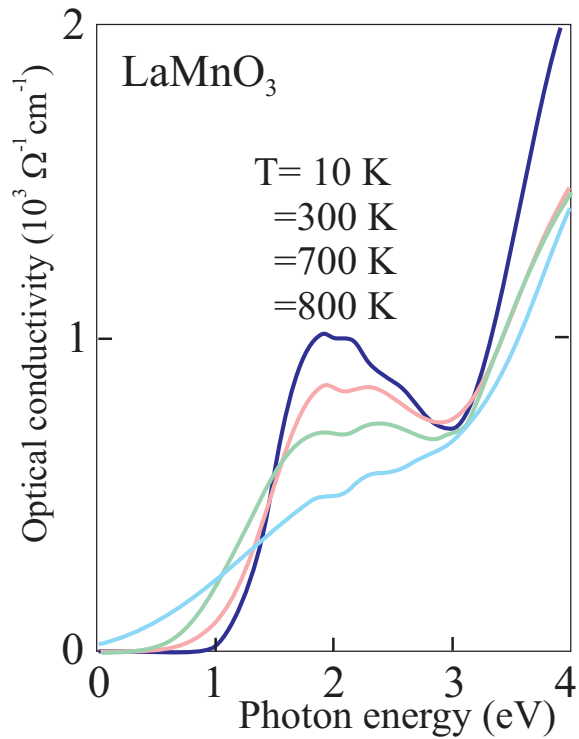}
\caption{(Color online) Temperature dependence of optical conductivity of parent LaMnO$_3$ for $\bf E\parallel ab$ (reproduced from Ref.\,\onlinecite{Tobe}).  } \label{fig7}
\end{figure}

Charge transfer instability and the CT exciton self-trapping in nominally pure manganites are indeed supported by the studies of their optical response. 

Anisotropic optical conductivity spectra  for a detwinned single crystal of LaMnO$_3$, which undergoes the orbital ordering below T$_{JT}\approx$\,780\,K has been derived from the reflectivity spectra  investigated   by Tobe {\it et al.}\,\cite{Tobe} over a wide temperature range, 10\,K$<$\,T\,$<$\,800\,K (see Fig.\,\ref{fig7}). As temperature is increased, the EH-dimers generating $d-d$ CT transition peaked around 2 eV shows a dramatic loss  of  spectral weight with its partial  transfer to the low energies. Simultaneously one observes a suppression of optical anisotropy. Above T$_{JT}$, the gap feature becomes obscure and the anisotropy disappears completely.
 Such a behavior of the 2 eV band can be hardly explained by the effect of spin fluctuations\,\cite{Kovaleva}, most likely it points to a shrinking of the A-AFI phase volume fraction with approaching to T$_{disp}$=T$_{JT}$ and phase transition to a novel unconventional metallic-like phase.   However, the optical conductivity does not reveal 
any signatures of Drude peak, that together with a rather large resistivity\,\cite{Good1}
points to an unusual charge transport. 
 
\begin{figure}[t]
 \includegraphics[width=8.5cm,angle=0]{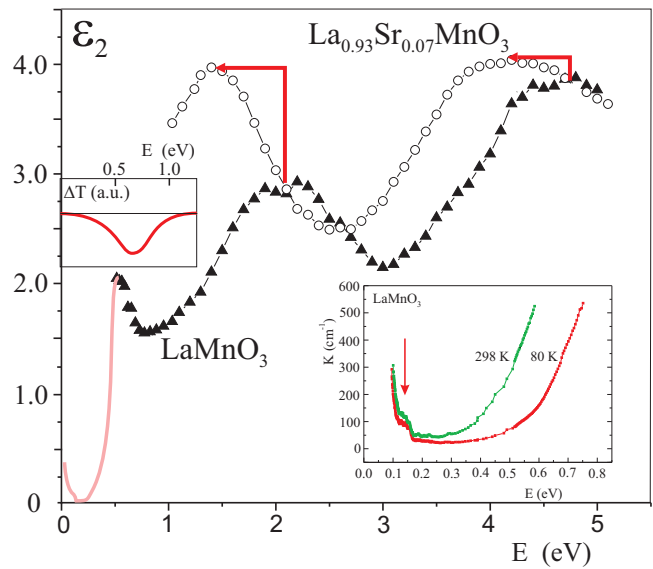}
 \caption{(Color online) Imaginary part of the dielectric function $\varepsilon _{ab}$ in LaMnO$_3$ (solid triangles) and La$_{0.93}$Sr$_{0.07}$MnO$_3$ (open circles), respectively\,\cite{optics-PS}. Low-energy part of the spectrum is guided on eye from the infrared absorption data (see right-hand inset). Right-hand inset: infrared absorption  for parent LaMnO$_3$ at 80 and 298\,K (reproduced from Ref.\,\onlinecite{optics-PS}). Left-hand inset: photoinduced transmittance of parent LaMnO$_3$ at T=25\,K (reproduced from Ref.\,\onlinecite{Mih})}
 \label{fig8}
 \end{figure}

 Main features of the optical response\,\cite{Tobe}  agree with predictions followed from the EPS phase diagram and isotropic character of the optical response of EH droplets. 
However, the reflectivity data did not reveal any midgap structures which observation and identification needs usually in direct absorption/transmission measurements. The most detailed studies of spectral, temperature and doping behavior of the midgap bands were performed in Refs.\,\onlinecite{optics-PS,optics-PSa,optics-PSb,droplet,CIC}.
All the manganites investigated, both parent and hole/electron doped, show up two specific low-energy optical features peaked near 0.10-0.15 eV (0.1 eV band) and 0.3-0.6 eV (0.5 eV band), respectively.  Results of the ellipsometric and direct absorption measurements for a single-crystaline parent LaMnO$_3$ sample are shown in Fig.\,\ref{fig8}, these directly reveal both 0.1 and 0.5 eV features in the spectral window of the bare matrix\,\cite{optics-PS}. These two bands can be naturally attributed to the CT transitions within the immobile EH-dimers and to the dimer transport activating transitions, respectively. Respective energies agree with theoretical predictions, although  more accurate value 0.15 eV for the "0.1 eV" peak points most likely to an essential electron-lattice effect.

The 0.5 eV band in LaMnO$_3$ was revealed by photoinduced absorption spectroscopy under light excitations with the photon energy near 2.4 eV that provides  optimal conditions for the EH pairs creation. Photoinduced   absorption was observed\,\cite{Mih} with a strong broad midinfrared peak centered at $\sim$ 5000 cm$^{-1}\approx 0.62$ eV. 
 Since the laser photoexcitation and measurement are pseudocontinuous, the photoexcited
EH-pair lifetimes need to be quite long for any significant photoexcited EH-pair density to build up. It means that the lattice is arranged in the appropriate relaxed state.
 The origin of the photoinduced (PI) absorption peak was  attributed\,\cite{Mih} to the photon-assisted hopping of anti-Jahn-Teller polarons formed by photoexcited charge carriers. This interpretation was based on the assumption of primary p-d CT transition induced by excitation light with the  energy $h\nu =2.41$ eV. However, the $d$-$d$ CT transition nature of 2 eV absorption band in LaMnO$_3$\,\cite{Kovaleva} unambiguously points to the EH dimers to be main contributor to  PI absorption peak. 
 In such a case, the PI absorption peak energy ($\sim 0.6$ eV) may be attributed to the energy of the photon-assisted hopping of the relaxed EH dimers (see Fig.~5) and can be used as an estimate of the Jahn-Teller energy $\Delta_{JT}$. 
 
Similar, so-called mid-gap features in nominally pure manganites were directly or indirectly observed by many authors. Furthermore, it seems that some authors  did not report the optical data below 1.5 eV to avoid the problems with these odd features. Observation of the MIR features agrees with the scenario of well developed intrinsic electronic inhomogeneity  inherent to nominally stoichiometric insulating manganites and composed of volume fraction of conceivably EH droplet phase.    

Finally, it is instructive to compare the midgap absorption spectrum of parent manganite with IR optical spectra of chemically doped compounds to see whether the nonisovalent substitution stimulates the condensation of EH-pairs and respective rise of the EH droplet volume fraction. Indeed, Okimoto {\it et al.}\,\cite{Okimoto} 
observed in  La$_{0.9}$Sr$_{0.1}$MnO$_3$ a broad absorption peaked around 0.5 eV which is absent at room temperature and increases in intensity with decreasing temperature.  In addition, the  absorption feature reported also shifts to lower energy as doping is increased, in agreement with  PI measurements\,\cite{Mih}. A mid-gap state with a similar peak energy and similar doping dependence was also observed at room temperature by Jung
{\it et al.}\,\cite{Jung}  in La$_{1-x}$Ca$_{x}$MnO$_3$.

Thus we see that the  strong and broad midinfrared optical feature  peaked near 0.5 eV and observed in all the perovskite manganites studied can be surely attributed to the  optical response of isolated EH-dimers or small EH droplets edged by the JT Mn$^{3+}$ centers, more precisely, to an optical activation of the dimer transport in such a surroundings. The peak energy may be used to estimate the Jahn-Teller splitting for $e_g$ levels in Mn$^{3+}$ centers and its variation under different conditions.

\subsection{Lattice effects in parent LaMnO$_3$} 

The unusual abrupt unit cell volume contraction by 0.36\%  has been observed by Chatterji {\it et al.}\,\cite{Chatterji} in LaMnO$_3$ at T$_{JT}$. The high-temperature phase just above T$_{JT}$ has less volume than the low-temperature phase.

The local structure of stoichiometric LaMnO$_3$ across the Jahn-Teller transition at T$_{JT}$ was studied by means of extended x-ray absorption fine structure (EXAFS) at Mn K-edge\,\cite{Sanchez} and high real space resolution atomic pair distribution function (PDF) analysis\,\cite{Qiu}. The both techniques reveal  two different Mn-O separations 1.92\,{\AA}(1.94\,{\AA}) and 2.13\,{\AA}(2.16\,{\AA}) distributed with intensity 2:1, respectively. Comparing these separations with room temperature neutron diffraction data\,\cite{Alonso}(1.907, 1.968, and 2.178\,{\AA})  the both groups point to  a persistence of the JT distortions of MnO$_6$  octahedra on crossing T$_{JT}$. However, both this result and that of Chatterji {\it et al.}\,\cite{Chatterji} most likely point to a transition to Mn-O separations specific for EH-dimers, or   nearest neighbor electron MnO$^{10-}_6$ (Mn$^{2+}$) and hole MnO$^{10-}_6$ (Mn$^{4+}$) centers coupled by fast electron exchange. Anycase the picture is that in the high-temperature O phase the local distortions of the Mn-O separations  are dynamical in character similar to those in BaBiO$_3$. 
 A signature of that is an excess increase of the thermal factors of oxygen atoms in going from O' to the O phase\,\cite{Rodriguez}.

The observed Raman spectra for undoped LaMnO$_3$ crystal at ambient pressure and room temperature reveal a number of additional lines, in particular, strong (A$_{1g}$+B$_{2g}$) mode 675 cm$^{-1}$, which are also have been observed in the spectra of doped  materials and may be attributed to droplets of EHBL phase\,\cite{Raman}.

Strong variation of the LaMnO$_3$ Raman spectra, both of intensity and energy shift with increasing laser power
\cite{Raman-1} could be related to the photo-induced nucleation and  the volume
expansion of the EH Bose-liquid. Surely, laser annealing can simply increase the temperature thus resulting in a A-AFI/EHBL volume fraction redistribution.
 The strong variations of the LaMnO$_3$ Raman spectra on the excitation laser power 
provide evidence for a structural instability that may result in a laser-irradiation-induced structural phase transition.
It is worth noting a strong  resonant character of the excitation of the Raman specta\,\cite{Raman-2} that points to a need in more extensive studies focused on the search of the EH droplet response.

The intrinsic electronic phase separation inherent for nominally undoped stoichiometric LaMnO$_3$ manifests itself in remarkable variations of X-ray diffraction pattern, optical reflectivity and Raman spectra, resistivity under pressures up to 40 GPa\,\cite{Loa}. 
The pressure induced variations of Raman spectra, in particular, a blue-shift and the intensity loss of the in-phase O2 stretching $B_{2g}$ mode with a concomittant  emergence of a new peak at $\sim$\,45 cm$^{-1}$ higher in energy evidenced some kind of electronic phase separation with a steep rise of the volume fraction   of the domains of a new phase within the parent A-AFI phase ("sluggish" transition\,\cite{Loa}). Evolution of new phase was accompanied by a dramatic change of reflectance which resembles that of for LaMnO$_3$ at ambient pressure on heating from low temperatures to T\,$>$\,T$_{JT}$\,\cite{Tobe}.  Furthermore, the system exhibited an anomalously strong pressure-induced fall of the room-temperature resistivity by three orders of magnitude in the range 0-30 GPa with an IM transition at 32 GPa. An overall fall of resistivity in the range 0-32 GPa   amounts  to five orders  of magnitude. However, the system retains a rather high resistance, exhibiting a "poor" metallic behavior typical for EHBL phase. It is worth  noting that at high pressures $>$\,30\,GPa the resistivity does not reveal sizeable temperature dependence between 80 and 300\,K  similarly to  the high-temperature T\,$>$\,T$_{JT}$ behavior of LaMnO$_3$ at ambient pressure (see Ref.\,\onlinecite{Good1} and Fig.\,1).
Overall these data provide a very strong support for our scenario of the A-AFI/EHBL electronic phase separation in parent manganite taking place without any hole/electron doping.

The effect of the O$^{16}$$\rightarrow$O$^{18}$ isotope substitution on the IM transition and optical response\,\cite{isotope} can be easily explained as a result of an energy stabilization of the parent A-type antiferromagnetic phase as compared with the EH Bose liquid. The percolation mechanism of the isotope effect in manganites is considered in Ref.\,\onlinecite{isotope}.

\subsection{Magnetic and resonance properties of EHBL phase in LaMnO$_3$} 
What about the magnetic properties of the novel phase?
In framework of our scenario the EH Bose liquid in LaMnO$_3$ evolves from the EH-dimers which are new peculiar magnetic centers with intrinsic spin structure and with enormously large magnetic moments in their ground ferromagnetic state. However, the EH dimers exist as  well defined entities only at very initial stage of the EHBL evolution. Within well developed EH Bose liquid we deal with a strong overlap of EH-dimers when these lose individuality. A tentative analysis of the EH liquid phase in parent manganites\,\cite{Moskvin-02} shows that it may be addressed to be a triplet bosonic analogue of a simple fermionic double exchange model with a well developed trend to a ferromagnetic  ordering. It is interesting that both models have much in common that hinders their discerning. In both cases the net magnetic moment of calcium(strontium)-doped manganite La$_{1-x}$Ca(Sr)$_x$MnO$_3$ saturates to the full ferromagnetic value  $\approx (4-x)\mu_B$ per formula unit. 
Well developed ferromagnetic fluctuations within EHBL  phase in LaMnO$_3$ have been observed in high-temperature susceptibility measurements by Zhou and  Goodenough\,\cite{Good1} which measured the temperature dependence of  paramagnetic susceptibility both below  and above T$_{JT}$. They observed  a change from an anisotropic antiferromagnetism to an isotropic ferromagnetism crossing T$_{JT}$ accompanied by an abrupt rise of magnetic susceptibility. These data point to an energy stabilization of the EH Bose liquid in an external magnetic field as compared with a parent A-type antiferromagnetic phase. 

The dc magnetic susceptibility shows two distinct regimes\,\cite{Causa,Tovar} for LaMnO$_3$, above and below T$_{JT}$. For  T\,$>$\,T$_{JT}$, $\chi_{dc}(T)$  follows  a  Curie-Weiss  (CW)  law, $\chi _{dc}(T)= C/(T - \Theta))$,  with  C =  3.4 $emu\cdot K/mol$ ($\mu _{eff}\approx 5.22 \mu _B$)  and $\Theta \approx 200$\,K. For  T\,$<$\,T$_{JT}$  the behavior of magnetic susceptibility strongly depends on the samples studied. Zhou and Goodenough\,\cite{Good1} observed  an abrupt fall  in the Weiss constant on crossing T$_{JT}$ from large ferromagnetic to a small antiferromagnetic $\Theta \approx 50$\,K, while Causa {\it et al.}\,\cite{Causa,Tovar} found that the  Curie-Weiss  behavior of $\chi_{dc}(T)$  is recovered only near room temperature with a reduced antiferromagnetic $\Theta \approx 75$\,K. Interestingly that instead of a natural suggestion of an electronic phase separated state below T$_{JT}$ with a coexistence of low- and high-temperature phases and steep change in effective $\Theta$, the authors\,\cite{Causa,Tovar} explained their data as a manifestation of dramatic changes in exchange parameters induced by crystal distortions. They refer to  theoretical calculations\,\cite{Solovyov} which show that J$_{ab}$ in parent manganites  is FM  and decreases with the JT distortion while J$_c$ changes from FM in the pseudocubic O phase 
to AFM in the O'-phase. However, the aforementioned estimations\,\cite{Ovanesyan} based on the experimental data for isostructural orthoferrites, orthochromites, and mixed  orthoferrites-chromites point to a more reasonable antiferromagnetic orbitally averaged exchange coupling of  two Mn$^{3+}$ ions with bond geometry typical for LaMnO$_3$: $J\approx 12.6$\,K.

Magnetic measurements for low-hole-doped LaMnO$_3$ samples\,\cite{Korolyov,Mukhin,Algarabel,Souza,Markovich} reveal a coexistence of antiferromagnetic matrix with   ferromagnetic clusters or spin glass behavior,  accompanied by a magnetic hysteresis phenomena. Anomalous magnitudes of the effective magnetic moment per manganese
 ion that considerably exceed expected theoretical values, up to $\mu _{eff}\approx 6\mu_B$ in La$_{0.9}$Sr$_{0.1}$MnO$_3$\,\cite{Mukhin} were explained to be an evidence of a disproportionation 2Mn$^{3+}$$\rightarrow$Mn$^{4+}$+Mn$^{2+}$\,\cite{Korolyov} or a superparamagnetic behavior of ferromagnetic clusters\,\cite{Mukhin}. 
As a whole, magnetic measurements for nearly stoichiometric LaMnO$_3$ support the disproportionation scenario.  
 
 The ESR spectrum of LaMnO$_3$ in a wide temperature range above T$_N$ and up to temperature $\sim$\,800\,K above T$_{JT}$ shows a single Lorentzian line with g\,$\sim$\,1.98-2.00 and $\Delta H\sim$\,2400  Gauss  at room  temperature\,\cite{Causa,Oseroff}. In common, the spectrum intensity follows the dc susceptibility,  however, the consistent interpretation of the origin of ESR signal, especially in O "pseudocubic" phase is still lacking. Two different electronic phases are documented by electronic spin resonance (ESR) measurements in slightly La-deficient  La$_{0.99}$MnO$_3$\,\cite{Markovich}. Further experimental ESR studies have to be carried out to clarify the issue.

The  ${}^{55}$Mn nuclear magnetic resonance (NMR) data support most likely the EHBL scenario. Indeed, the zero-field ${}^{55}$Mn NMR spectrum  in a nominally undoped LaMnO$_3$ consists of a sharp central peak at 350 MHz due to bare Mn$^{3+}$O$_6^{9-}$ centers and two minority signals at approximately 310 and 385 MHz\,\cite{Allodi},  that can be assigned to a localized hole MnO$_6^{8-}$ (=Mn$^{4+}$) center and EH-dimers with a fast bosonic exchange, respectively. Evolution of such a picture with Ca(Sr) doping can easily explain a complex ${}^{55}$Mn NMR lineshape in  La$_{1-x}$Ca(Sr)$_x$MnO$_3$ samples\,\cite{Algarabel,Allodi}. It is worth noting that Tomka {\it et al.}\,\cite{Tomka} observed three ${}^{55}$Mn NMR signals in a hole doped PrMnO$_3$ around 310, 400 and 590 MHz, which can be attributed to localized hole MnO$_6^{8-}$ and electron  MnO$_6^{10-}$ centers (narrow resonances around 310  and 590 MHz, respectively) and to EH droplets with a fast bosonic exchange (broad resonance around 400 MHz).

Ii is worth noting that the ${}^{55}$Mn NMR lineshape in  La$_{1-x}$Ca(Sr)$_x$MnO$_3$ samples\,\cite{Allodi,Algarabel} with a most part of intensity shifted to a very broad line in the range 350-450 MHz can hardly be explained in framework of a so-called DE (double exchange) line\,\cite{Allodi} with a frequency $f_{DE}=\frac{1}{2}(f(Mn^{3+})+f(Mn^{4+}))$ derived from that typical for Mn$^{3+}$(350 MHz) and Mn$^{4+}$ (310 MHz). Our scenario with a broad line centered with more or less redshift from a frequency specific for a high-spin state of the EH-dimer: $f_{EH}=\frac{1}{2}(f(Mn^{2+})+f(Mn^{4+}))\approx 450$ MHz with $f(Mn^{2+})\approx 590$ MHz and $f(Mn^{4+})\approx 310$ MHz is believed to be more appropriate one.
It is worth noting that the ${}^{55}$Mn NMR response of EH-dimers can shed some light  on several  ${}^{55}$Mn NMR puzzles, in particular, observation of the low-temperature (4.2\,K) low-frequency  NMR lines at 260 MHz in one of nominally undoped LaMnO$_3$ samples\,\cite{Allodi-1} and even  at 100 MHz in  a more complex manganite (BiCa)MnO$_3$\,\cite{Shimizu}. In both cases we deal seemingly with a some sort of a stabilization of low-spin states for EH-dimers, for instance, due to the Mn-O-Mn bond geometry distortions resulting in an antiferromagnetic Mn$^{2+}$-O-Mn$^{4+}$ superexchange.

The ${}^{55}$Mn NMR spectra of slightly nonstoichiometric LaMnO$_3$\,\cite{Kapusta} may be viewed as the most striking evidence of the EH dimer response in a spin inhomogeneous glass-like state. A simple comparison of experimental spectra with theoretical predictions for EH dimers (see Fig.\,\ref{fig4}) shows a clear manifestation of the S=4,3,2 spin multiplets of the EH dimers with the mixing effects due to a spin noncollinearity.

Magnetic and transport properties of a single-crystalline parent undoped manganite LaMnO$_3$ have been studied recently under  ultra-high megagauss  magnetic field at helium temperatures\,\cite{Kudasov}.  In accordance with theoretical predictions\,\cite{Bolzoni} a sharp magnetic spin-flip transition was observed at about 70 T without visible transport anomalies.  On further rising the magnetic field the authors observed unusual magneto-induced IM transition at H$_{IM}\sim$\,220\,Tesla that is considerably above the field of the magnetic saturation of the A-AFI phase.    Large values of the  $p-d$ or $d-d$ charge transfer energies in bare A-AFI phase of parent manganites  ($\sim 2$\,eV in LaMnO$_3$\,\cite{Kovaleva})  makes the energy difference between the A-AFI ground state and any metallic phase seemingly too large to be overcomed even for  magnetic fields as large as hundreds of Tesla. Zeeman energy associated with such a field is clearly more than an order of magnitide smaller than the charge reodering energy. Thus we see that a puzzling field-driven IM transition  cannot be explained within a standard scenario implying the parent manganite LaMnO$_3$ to be an uniform system of the Jahn-Teller Mn$^{3+}$ centers with an A-type antiferromagnetic order and needs in a revisit of our view on the stability of  its ground state.
However, our scenario  can easily explain the puzzling field driven IM transition in perovskite manganite LaMnO$_3$\,\cite{Kudasov} to be a result of a percolative transition in an inhomogeneous phase-separated A-AFI/EHBL state. 
The volume fraction of the  ferromagnetic EHBL phase grows in an applied magnetic field, and at a sufficiently high field this fraction reaches its percolation threshold to give the IM transition. 
It is clear that a relatively small zero-field volume fraction of novel ferromagnetic EHBL phase in the parent manganite have required large magnetic field to induce the IM transition.

\subsection{Dielectric anomalies in LaMnO$_3$}
The broadband dielectric spectroscopy helps in characterizing the phase states and transitions in  Mott insulator. Above we pointed to anomalous electric polarisability of the EH dimers and EH droplets that would result in dielectric anomalies in the EHBL phase and the phase-separated state of LaMnO$_3$. Indeed, such anomalies were reported recently both for poly- and single-crystalline samples of parent LaMnO$_3$\cite{Mondal}. First of all, one should note relatively high static dielectric constant in  LaMnO$_3$ at T\,=\,0 ($\varepsilon_0\sim 18-20$)  approaching to values typical for genuine multiferroic systems  ($\varepsilon_0 \approx 25$), whereas for the conventional nonpolar systems,  $\varepsilon_0$ varies within 1-5.
The entire $\varepsilon^{\prime}(\omega,T)$- T pattern across 77-900\,T has
two prominent features: (i)  near T$_N$ and  (ii)  near T$_{JT}$ to be essential signatures of puzzlingly unexpected multiferroicity. 
Far below T$_N$,
$\varepsilon^{\prime}(\omega,T)$   is nearly temperature and frequency independent, as expected. Following the anomaly at T$_N$, $\varepsilon^{\prime}(\omega,T)$  rises with T by 5 orders of magnitude near T$_{JT}$. Finally, $\varepsilon^{\prime}$ becomes nearly temperature independent beyond T$_{JT}$. 
The P-E loop  does not signify any ferroelectric order yet the time-dependence plot resembles the "domain-switching-like" pattern. The finite loop area signifies the presence of irreversible local domain fluctuations. From these results, it appears that the intrinsic  electrical polarization  probably develops locally with no global ferroelectric order.

The nature of the anomaly at T$_{JT}$ varies with the increase in Mn$^{4+}$ concentration following a certain trend — from a sharp upward feature to a smeared plateau
and then a downward feature to finally, a rather broader downward peak.  

The observation of an intrinsic dielectric response  in globally centrosymmetric LaMnO$_3$, where no ferroelectric order is possible due to the absence of off-centre distortion in MnO$_6$ octahedra cannot be explained in frames of the conventional uniform antiferromagnetic insulating  A-AFI scenario and agrees with the electronic A-AFI/EHBL phase separated state with a coexistence of non-polar A-AFI phase and highly polarisable EHBL phase.

\subsection{Comment on the experimental non-observance of new phase in LaMnO$_3$}

By now there has been no systematic exploration of exact valence and spin
state of Mn in  perovskite manganites. Using electron paramagnetic resonance (EPR)
measurements Oseroff {\it et al}.\,\cite{Oseroff}
 suggested that below 600 K in LaMnO$_3$ there are no isolated Mn atoms of valence $+2,+3,+4$, however they  argued that EPR signals are consistent with a complex magnetic entity composed of Mn ions of different valence.

Park et al. \cite{Park} attempted to support the  Mn$^{3+}$/Mn$^{4+}$ model,
based on the Mn $2p$ x-ray photoelectron spectroscopy (XPES) and $O1s$
absorption. However, the significant discrepancy between the weighted
Mn$^{3+}$/Mn$^{4+}$ spectrum and the experimental one for given
 $x$ suggests a more complex doping effect. Subias et al.\cite{Subias} examined the valence
 state of Mn utilizing Mn $K$-edge x-ray  absorption near edge spectra (XANES), however,
  a large discrepancy is found between experimental spectra given intermediate doping and
  appropriate superposition of the end members.

The valence state of Mn in Ca-doped LaMnO$_3$ was studied by high-resolution
Mn $K\beta $ emission spectroscopy by Tyson {\it et al.} \cite{Tyson}. No
evidence for Mn$^{2+}$ was claimed at any $x$ values seemingly ruling out
proposals regarding  the Mn$^{3+}$ disproportionation. However, this conclusion
seems to be absolutely unreasonable one. Indeed, electron center
 MnO$_{6}^{10-}$ can be found in two configuration with formal Mn valence Mn$^{2+}$ and
 Mn$^{1+}$ (not simple Mn$^{2+}$), respectively. In its turn, the hole center MnO$_{6}^{8-}$
 can be found in two configuration with formal Mn valence Mn$^{4+}$ and Mn$^{3+}$ (not simple Mn$^{4+}$), respectively. Furthermore, even the bare center MnO$_{6}^{9-}$
 can be found in two configuration with formal Mn valence Mn$^{3+}$ and Mn$^{2+}$ (not simple Mn$^{3+}$), respectively. So, within the model the Mn $K\beta $ emission spectrum for the Ca-doped LaMnO$_3$ has to be a superposition of appropriately weighted Mn$^{1+}$, Mn$^{2+}$, Mn$^{3+}$, and Mn$^{4+}$ contributions (not simple Mn$^{4+}$ and Mn$^{3+}$, as one assumes in Ref.\,\onlinecite{Tyson}). Unfortunately, we do not know the Mn $K\beta $ emission spectra for the oxide
   compounds  with Mn$^{1+}$ ions, however a close inspection of the Mn $K\beta $ emission
   spectra for the series of  Mn oxide compounds with Mn valence varying from $2+$ to $7+$
   (Fig.\,2 in Ref.\,\onlinecite{Tyson}) allows to uncover a rather clear dependence on valence, and
   indicates a possibility to explain the experimental spectrum for Ca-doped LaMnO$_3$ ($ibid$, Fig.\,4a)
    as a superposition of appropriately weighted Mn$^{1+}$, Mn$^{2+}$,
   Mn$^{3+}$, and Mn$^{4+}$
    contributions. 
Later\,\cite{Gilbert} it has been shown that Mn $L$ edge absorption rather than that of  $K$ edge is completely dominated by Mn 3d states and, hence, is an excellent indicator
of Mn oxidation state and coordination. Interestingly that the results of the X-ray absorption and emission  spectroscopy in vicinity of the Mn $L_{23}$ edge\,\cite{Jimenez} provide a striking evidence of a coexistence of Mn$^{3+}$ and Mn$^{2+}$ valent states in a single crystalline LaMnO$_3$.

This set of conflicting data together with a number of additional data
\cite{Croft} suggests the need for an in-depth exploration of the Mn valence
problem  in this perovskite system. However, one might say, the doped
manganites are not only systems with   mixed valence, but systems with
indefinite valence, where we cannot, strictly speaking,  unambiguously
distinguish Mn species with either distinct valence state.

It seems, by now there are no techniques capable of direct and unambiguous detection of  new  electron-hole Bose liquid. However, we do not see any sound objections against such a scenario that is shown to explain a main body of experimental data.


\section{Hole doping of parent manganite}

Evolution of the electronic structure of nominally insulating 3d oxides under a nonisovalent substitution as in La$^{3+}_{1-x}$Sr$^{2+}_x$MnO$_3$ remains one of the challenging problems in  physics of strong correlations. A conventional model approach focuses upon  a hole doping and implies a change in the (quasi)particle occupation in the valent band or a hole localization in either cation 3d orbital or anion O 2p orbital, or in a proper hybridized molecular orbital. However, in the 3d oxides unstable with regard to a charge transfer such as parent manganites one should expect just another scenario when the nonisovalent substituents do form the nucleation centers for the EH droplets thus provoking the first order phase transition into a novel EH disproportionated phase with a proper deviation from a half-filling.

Conventional double exchange model implies the manganese location of the doped hole and its motion in the lattice formed by nominal parent manganite\,\cite{Dagotto}. However, by now there are very strong hints at oxygen location of doped holes. One might point to
several exciting  experimental results supporting the oxygen nature of holes in manganites. The first is a direct observation of the O 2p holes in the O 1s x-ray absorption spectroscopy measurements\,\cite{Ju}. Second, Tyson {\it et al.}\,\cite{Tyson} in their Mn $K\beta $ emission    spectra  studies of the Ca-doped LaMnO$_3$ have observed an "arrested" Mn-valence response to the doping in the $x<0.3$ range, also consistent with creation of predominantly oxygen holes. Third, Galakhov {\it et al.}\,\cite{Galakhov} have reported Mn 3s x-ray emission spectra in mixed-valence manganites and shown that
the change in the Mn formal valency from 3  to 3.3  is not
accompanied by any decrease in the Mn 3s splitting. They have proposed that this effect can be explained by the appearance in the  ground-state configuration of holes in the O 2p states. The oxygen location of the doped holes is partially supported by   observation of
anomalously large magnitude of saturated magnetic moments in ferromagnetic
state for different doped manganites \cite{Korolyov,Mukhin}.

 Two oxygen-hole scenarios are possible. The first implies the hole doping directly to bare   A-AFI phase of parent manganite. Given light doping we arrive at the hole trapping in potential wells created by the substituents such as Ca$^{2+}$, Sr$^{2+}$ or cation vacancies.  This gives rise to evolution of hole-rich, orbitally disordered ferromagnetic phase.  The volume fraction of this phase increases with x, and ferromagnetic ordering within this phase introduces spin-glass behavior where the ferromagnetic phase does not percolate in zero magnetic field  H=0; but growth of the ferromagnetic phase
to beyond percolation in a modest field can convert the spin glass to a bulk ferromagnetic insulator. On further increasing the hole doping the ferromagnetic metallic ground state  is obtained with  itinerant oxygen holes and degenerate e$_g$ orbitals of Mn$^{3+}$ ions. 

In  second scenario one proposes that doped holes trigger the phase transition to an "asymmetrically" disproportionated phase with nominal non-JT Mn$^{2+}$ ions and oxygen holes that can form a band of itinerant carriers. This scenario implies that the doped holes simply change a hole band filling.

The both scenario imply an  unconventional system with two, Mn 3$d$ and O $2p$, unfilled shells. One should note that
despite a wide-spread opinion the correlation effects for the oxygen holes can be
rather strong. These could provide a coexistence  of the  two (manganese and oxygen) nonfilled bands.

Such a $p-d$ model with ferromagnetic $p-d$ coupling immediately
explains many unconventional properties of the hole doped manganites. First of
all, at low hole content we deal with hole localization in impurity potential.
Then, given further hole doping a percolation threshold occurs accompanied by
insulator-anionic oxygen metal phase transition and ferromagnetic ordering both
in oxygen and Mn sublattices, due to  a strong ferromagnetic Heisenberg $pd$
 exchange. However, it should be noted that ferromagnetic sign of $pd$ exchange
 is characteristic of nonbonding $p$ and $d$ orbitals.

The oxygen hole doping results in a strong spectral weight transfer from the
intense O$2p$-Mn$3d$  CT transition    bands to the O$2p$ band developed. The
Mn$^{3+}$ $d-d$ transitions will gradually shift to the low energies due to a
partial O$2p$ hole screening of the crystalline field. In a whole, optical data
do not disprove the oxygen hole scenario. 


Despite many controversial  opinions regarding the electronic structure of doped holes the 
current description of complex phase diagrams for doped manganites implies a well-developed phase separation with coexistence of bare antiferromagnetic and several ferromagnetic phases\,\cite{Dagotto,Good-04}. What is the role played by the EHBL phase inherent for parent manganites? 

Hole doping of parent manganite is produced by a nonisovalent substitution as in La$_{1-x}$Sr$_x$MnO$_3$ or by an oxygen nonstoichiometry. The Sr$^{2+}$,  Ca$^{2+}$ substituents form effective trapping centers for the EH dimers and the nucleation centers for the EH Bose liquid. At a critical substituent concentration $x_c\approx 0.16$  one arrives at a percolation threshold\,\cite{Gorkov} when  the conditions for an itinerant particle hopping do emerge.  
Holes are doped into EH Bose liquid of parent LaMnO$_3$ similar to generic BaBiO$_3$ system only pairwise, transforming formally electron MnO$_6^{10-}$ center to hole  MnO$_6^{8-}$ center. Similarly to BaBiO$_3$ doped hole centers form local composite bosons which shift the system from half-filling ($n_B=1/2$).

It seems the EHBL phase addressed above  appears to be an important precursor for a ferromagnetic metallic phase responsible for colossal magnetoresistance observed in doped manganites. Existence of such an intermediate "poor metallic" phase seems to be   essential for a transformation of bare insulating A-AFI phase to a "good-metallic" phase under hole doping. Low-energy CT excitations typical for EHBL phase and well  exhibited in optical response (see Figs.\,7,8) give rise to a significant screening of electrostatic interactions and to a suppression of localization trend for doped charge carriers with their escape out of charge traps and the evolution of itineracy. This trend is well illustrated in Fig.\,\ref{fig8}, where the dielectric function $\varepsilon_2$ is shown both for parent and slightly hole doped LaMnO$_3$. We see a clear red-shift both for low-energy (2 eV)  $d-d$ CT  band and high-energy (4.5 eV) $p-d$ CT  band with a  rise of intensity for both bands, particularly sharp for the 2 eV band. All these effects evidence the lowering of effective values for the charge transfer energies, that is a clear trend to "metallicity".

One of the intriguing issues is related with  seemingly masked superconducting fluctuations in doped manganites and its relation to colossal magnetoresistance.
Indeed, doped  manganites reveal many properties typical for
superconducting materials or, rather, unconventional superconductors such as cuprates.  Kim\,\cite{Kim} has proposed the frustrated p-wave pairing superconducting state similar to the A$_1$ state in superfluid He-3 to explain CMR, the sharp drop of resistivity, the steep jump of specific heat, and the gap opening in tunneling of manganese oxides. 
 In this scenario, colossal magnetoresistance (CMR) is naturally explained by the superconducting fluctuation with increasing magnetic fields. This idea is closely related to the observation of anomalous proximity effect between superconducting YBCO and a  manganese oxide, La$_{1-x}$Ca$_x$MnO$_3$  or La$_{1-x}$Sr$_x$MnO$_3$\,\cite{Kasai}, and also the concept of local superconductivity manifested by doped manganites\,\cite{Mitin}.

\section{Conclusion} 

To summarize, we  do assign  anomalous properties of parent   manganite LaMnO$_3$ to charge transfer instabilities and competition between insulating A-AFM phase and metallic-like dynamically disproportionated phase formally separated by a first-order phase transition at T$_{disp}$\,=\,T$_{JT}$$\approx$\,750\,K.  
We report a comprehensive elaboration of a so-called "disproportionation" scenario in manganites which was addressed earlier by many authors, however, by now it was not properly developed. The unconventional high-temperature phase is addressed to be a specific electron-hole Bose liquid  rather than a simple "chemically" disproportionated R(Mn$^{2+}$Mn$^{4+})$O$_3$ phase. 
We arrive at highly frustrated system of triplet $(e_g^2){}^3A_{2g}$ bosons moving in a lattice formed by hole Mn$^{4+}$ centers when the latter  tend to order G-type antiferromagnetically, the triplet bosons tend to order ferromagnetically both with respect to its own site and its nearest neighbors, nearest neighboring bosons strongly prefer an antiferromagnetic ordering. Lastly, the boson transport prefers an overall ferromagnetic ordering. 

Starting with different experimental data we have reproduced a typical  temperature dependence of the volume fraction of the high-temperature mixed-valent EHBL phase. 
New phase nucleates as a result of the CT instability and evolves from the self-trapped CT excitons, or specific EH-dimers, which seem to be a precursor of both insulating and metallic-like ferromagnetic phases observed in manganites. We present a detailed analysis of electronic structure, energy spectrum, optical, magnetic and resonance properties of EH-dimers.
 We argue that a slight nonisovalent substitution, photo-irradiation, external pressure or magnetic field give rise to an electronic phase separation with a nucleation or an overgrowth of EH-droplets. Such a scenario provides a comprehensive explanation of numerous puzzling  properties observed  in parent and nonisovalently doped  manganite LaMnO$_3$ including an intriguing manifestation of superconducting fluctuations.
 
 We argue that  the  unusual  ${}^{55}$Mn NMR spectra of nonisovalently doped  manganites LaMnO$_3$ may be addressed to be a clear signature of a quantum disproportionation and formation of EH-dimers.
 Given the complex phase separation diagram of this class of materials, the study of the nominally stoichiometric parent compound could give a deep insight into the physics governing the doped version of these manganese oxides. It would be important to verify the expectations of EHBL scenario by more extensive and goaled  studies.

I thank N.\,N. Loshkareva, Yu.\,P. Sukhorukov, K.\,N. Mikhalev,  Yu.\,B. Kudasov and V.\,V. Platonov  for stimulating and helpful discussions. The work was  supported by RFBR grants Nos.  06-02-17242,  07-02-96047,  and  08-02-00633.

\end{document}